\documentclass[%
reprint,
superscriptaddress,
amsmath,amssymb,
prb,
floatfix]{revtex4-2}
\usepackage{amsmath}
\usepackage{amsthm,wrapfig,pgf,tikz,pgfplots}
\usepackage[export]{adjustbox}
\usepackage{graphicx}% Include figure files
\usepackage[permil]{overpic}
\usepackage{dcolumn}% Align table columns on decimal point
\usepackage{bm}% bold math
\usepackage{hyperref}% add hypertext capabilities
\usepackage[normalem]{ulem}

\usepackage{calligra}

\DeclareMathAlphabet{\mathcalligra}{T1}{calligra}{m}{n}
\DeclareFontShape{T1}{calligra}{m}{n}{<->s*[2.2]callig15}{}

\pgfplotsset{compat=1.15}
\usetikzlibrary{arrows,calc,fadings,decorations.pathreplacing}

\begin{document}

%\preprint{APS/123-QED}

\title{Probing nonlocal correlations in magnetic rare-earth clusters}% 

\author{David W. Facemyer}
 \email{df008219@ohio.edu}
\affiliation{Department of Physics and Astronomy and Nanoscale and Quantum Phenomena Institute, Ohio University, Athens, OH 45701}%

\author{Sergio E. Ulloa}
% \email{ulloa@ohio.edu}
\affiliation{Department of Physics and Astronomy and Nanoscale and Quantum Phenomena Institute, Ohio University, Athens, OH 45701}%

\date{\today}% 

\begin{abstract}
Understanding and quantifying entanglement entropy is crucial to characterize the  quantum behaviors that drive phenomena in a variety of systems. Rare-earth spin complexes, with their unique magnetic properties, provide fertile ground for exploring these nonlocal correlations. In this work, we study Eu$^{2+}$ ions deposited on a Au(111) substrate, modeling finite clusters of large spin-moments using a Heisenberg Hamiltonian parameterized by first-principles calculations. Our analysis reveals a one-to-one correspondence between structures in the differential conductance profiles and changes in the von Neumann entanglement entropy of bipartite subsystems, influenced by probe-ion separation and applied magnetic fields. Distinct braiding patterns in the conductance profiles are shown to correspond to stepwise changes in the entanglement entropy, providing a new avenue for investigating quantum correlations. These results establish a foundation for experimentally probing and controlling entanglement in lanthanide-based systems, with potential applications in quantum technologies.
\end{abstract}

\maketitle

%\tableofcontents
%%%%%%%%%%%%%%%%%%%%%%%%%%%%%%%%%%%%%%%%%%%%
\section{\label{Introduction}Introduction}
%-----An entanglement backdrop-----
Entanglement, a many-body correlation that is not typically present in the classical nature of macroscopic bodies \cite{trimmer1980}, has attracted a great deal of attention lately in a diversity of quantum systems \cite{bell1964, pouranvari2023, laflorencie2016}. This phenomenon, where a quantum-mechanical state cannot be represented as a factorizable product state, is a defining characteristic of quantum mechanics.

%-----What is it good for?-----
Over the last few decades, entanglement is increasingly seen as a resource to be exploited in a new age of quantum data, otherwise inaccessible by classical means \cite{raussendorf2001, horodecki2009, zhang2024, hotta2009, rodriguez-briones2023}. A seminal paper on quantum teleportation by Bennett et al.\ strongly motivated this view \cite{bennett1993}. Every year since has ushered novel methods for manipulating quantum systems in unprecedented fashion, which in turn has sparked a resurgence in exploring entanglement further \cite{guhne2009}. Dense coding, for example, was recently realized using Einstein-Podolsky-Rosen (EPR) pairs, where up to two bits of classical information were transmitted using a single qubit \cite{piveteau2022}. And, being necessary for quantum information and sensing applications, quantum entanglement has recently drawn new attention in the quantum engine domain \cite{zhang2024}.

%-----Why is it difficult to experimentally measure?-----
Unfortunately, entanglement is not resilient to environmental conditions \cite{horodecki2009}, as experimental systems, no matter how sequestered, often include degrees of freedom that interfere with  environmental subsystems. This is also limiting for  local observations communicated classically (LOCC), which is one of the processes for communicating quantum data \cite{chitambar2014}. Discovering better ways of creating and manipulating systems that possess entanglement and sufficiently tolerate probe and manipulation techniques is now a paramount task for achieving stable and reliable quantum technologies.

%-----STM interest in spin systems-----
An experimental method that allows atomically precise quantum system engineering is scanning tunneling microscopy (STM) \cite{eigler1990, hla2005}. This technique is capable of elucidating  electronic and magnetic properties of surfaces and adatoms by examining signatures in the differential conductance and the coherent scattering of electrons \cite{ternes2015}. If such techniques could be used to characterize the degree of entanglement between systems, the STM would prove to be an even more formidable quantum tool. In this work, we will show that STM indeed offers a great unique approach to assess entanglement in adatom systems like those we consider.

%-----Rare Earth-----
Rare-earth (RE) elements offer an inherent advantage, as their $4f$ electrons are well shielded from environmental fluctuations \cite{zhou2020}. In addition, their emission wavelengths are compatible with current fiber optic infrastructure, which make them viable prospects for quantum communication \cite{xia2022}. Recently, researchers have shown high coherence times and qudit gate fidelities in novel electron-nuclear systems with large-spin \cite{fuentes2024}. They suggest that spin species whose Hilbert space spans many dimensions could offer richer means for error corrections. We have also discussed that certain RE atoms might serve as suitable candidates for systems with high spin and corresponding large Hilbert space \cite{facemyer2023}.

%-----Realistic system with DFT j-couplings-----
Single-ion anisotropy (SIA) and indirect exchange couplings between Eu adatoms, situated on the Au(111) surface, were extracted from density functional calculations. The adatoms are found in a $+2$ ionic state, which gives them a large spin-$\frac{7}{2}$ and corresponding magnetic moment. We construct a generalized exchange function, parameterized as a Ruderman-Kittel-Kasuya-Yosida (RKKY) interaction \cite{ruderman1954}, and model finite adatom clusters using the Heisenberg spin Hamiltonian \cite{choi2019}. Exact diagonalization is employed to evaluate the low-lying magnetic excitation spectrum, enabling analysis of the STM spectroscopy of the system. An explicit correspondence emerges between differential conductance profiles and the calculated von Neumann entanglement entropies as functions of applied external magnetic fields and/or separations between subsystems. This allows us to assess nonlocal correlations in these systems using a direct local probe (the STM tip), without the need for accurate but more elaborate techniques like quantum interference of twin systems \cite{islam2015}.

%-----Close the intro by summarizing main objectives of the research and potential implications-----
The theoretical analysis of differential conductance profiles reveals a direct correlation with the von Neumann entanglement entropy in these RE spin systems. The effective subsystem couplings, derived from RKKY interactions, are shown to produce the oscillatory behavior seen in the entanglement entropy as a probe-ion changes positions. Notably, the entanglement entropy shows clear plateaus corresponding to stable ground state configurations over regions of magnetic field, while regions of rapid change reflect transitions between magnetic ground states. These transitions are  evident when the applied magnetic field is perpendicular to the SIA, and intercrossing (`braiding') structures in the conductance profiles align with the plateaus in the entanglement entropy.  We propose that comparing these theoretical results with experimentally accessible STM data could provide a novel framework for assessing quantum entanglement.  Deeper understanding of this phenomenon could offer valuable insights into the development of entanglement in real quantum systems and help create more robust quantum technologies.
%%%%%%%%%%%%%%%%%%%%%%%%%%%%%%%%%%%%%%%%%%%%

\section{\label{Methods}Methods}
\begin{figure}
    \begin{overpic}[width=0.50\columnwidth, left]{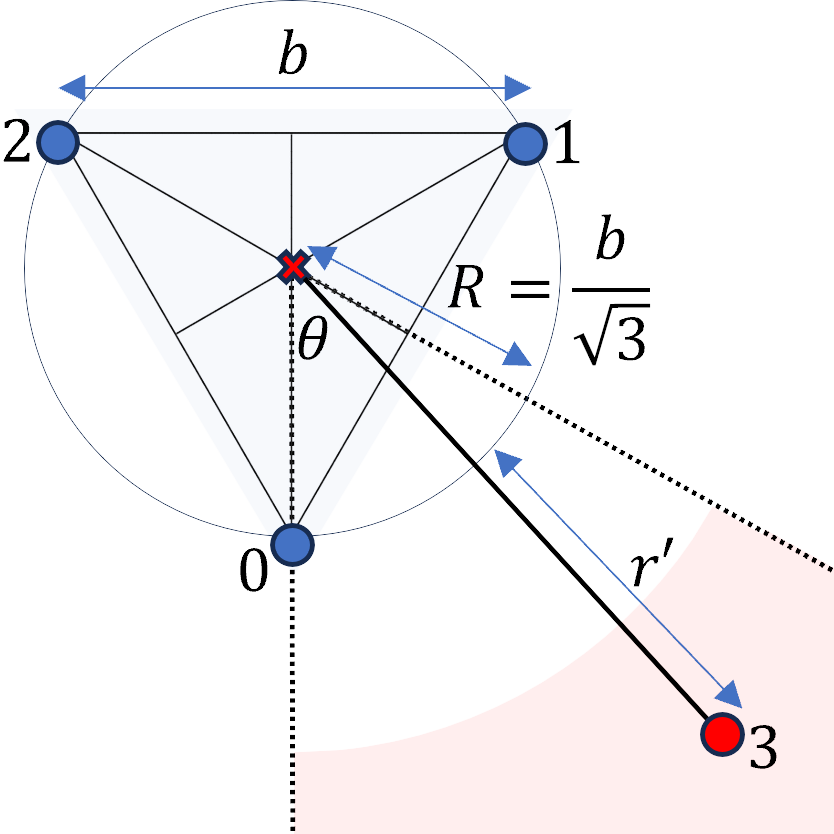}%
        \put(0,0){\includegraphics[width=0.50\columnwidth, right]{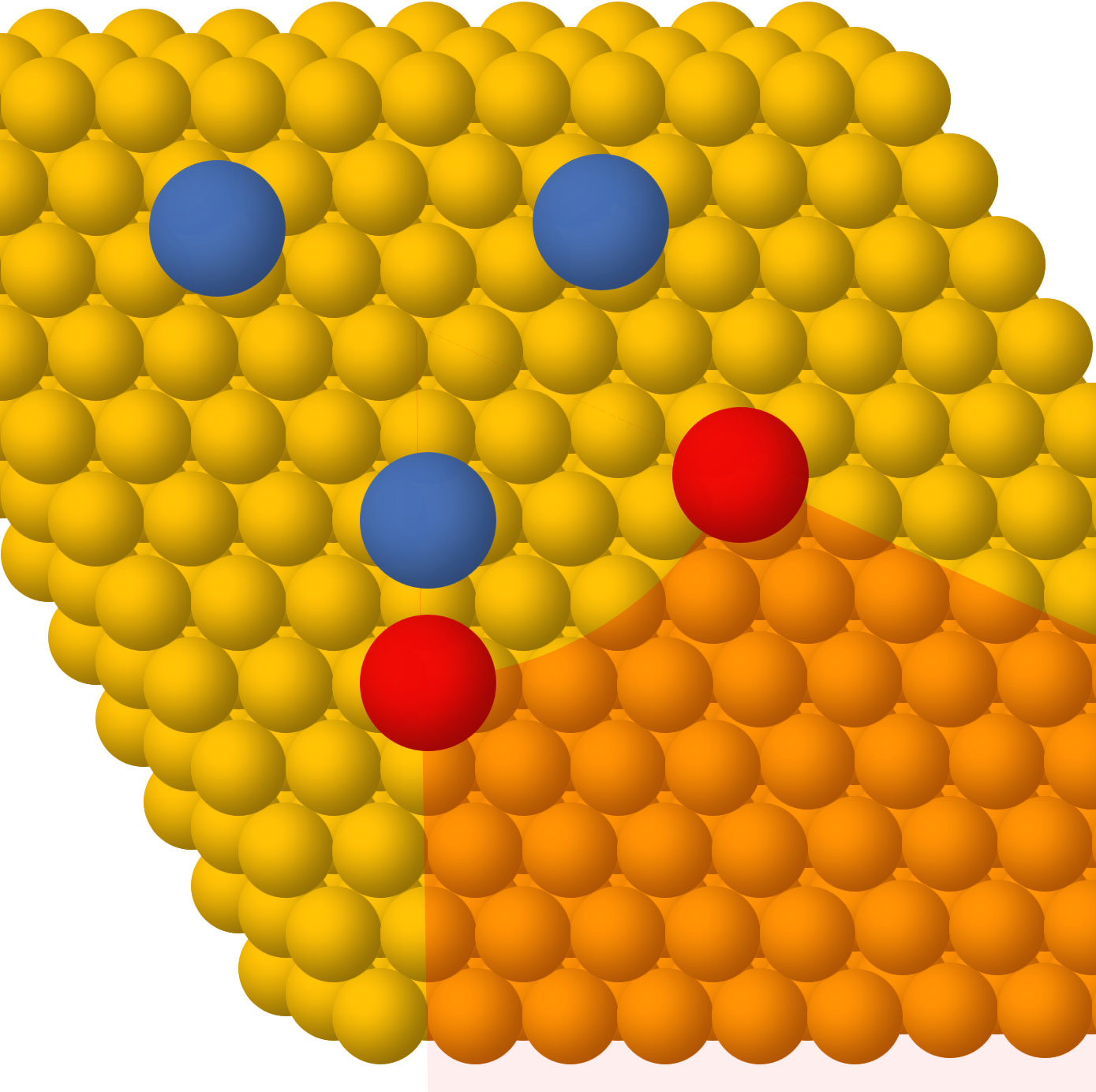}}%
        \end{overpic}
    \caption{An illustration of the composite bipartite system we consider. The blue-highlighted, equilateral triangular region of side $b$ represents the Eu$^{2+}$ ions $\{0,1,2\}$ (blue) comprising subsystem A, while subsystem B has only ion $\{3\}$ (red). We consider placement of ion $\{3\}$ in the region shaded red, at a distance $r'$ from the inner circle of radius $R={b/ \sqrt{3}}$. The sector bound by $0^{\circ} \leq \theta \leq 60^{\circ}$, where $\theta$ is measured with respect to the vertical line, repeats symmetrically for $\theta > 60^\circ$. The figure on the right shows two possible positions of the probe ion (red), $(r', \theta) \approx (5 \,\mathrm{\AA},0^{\circ})$, and $(5 \,\mathrm{\AA},60^{\circ})$, with all adatoms on bridge-sites on the Au(111) surface.}
    \label{system}
\end{figure}

We examine several finite cluster geometries constructed as two subsystems with varying separation and orientation. Figure \ref{system} illustrates such a bipartite/composite system as a blue-shaded triangular cluster representing subsystem A, with ions labeled $\{0, 1, 2\}$, while the probe-ion, labeled $\{3\}$, represents subsystem B. We focus on two different configurations: 1) subsystem A is arranged so that all the ions on the corners are ferromagnetically coupled to one another; and 2) a classically frustrated cluster where all corner ions interact antiferromagnetically with each other. These two configurations can be arranged by controling the inter-ion separation $b$ (see below). Energetics obtained from DFT suggests that the Eu$^{2+}$ ions we consider favor bridge-sites on the Au(111) surface, which results in a hexagonal grid of possible adsorption sites \cite{facemyer2023}. Such cluster systems can be assembled using atomic manipulation and probed by differential conduction under applied fields to explore the energetics and overall structure of the states \cite{hla2005, khajetoorians2019, kamlapure2018, figgins2019}.

The magnetic properties of these composite systems are well described by a Heisenberg Hamiltonian with full exchange tensor between species $i$ and $j$ \cite{choi2019},
\begin{equation}\label{FullHam}
    \hat{H} = \sum_{i<j}{\hat{\vec{S}}_i \cdot \mathrm{J}_{ij} \cdot \hat{\vec{S}}_j} \; .
\end{equation}
Similar adatom systems have been experimentally shown to interact predominately via itinerant electrons and possess significant SIA \cite{schuh2012}. The effective neighbor exchange coupling between adatoms $i$ and $j$ separated by a distance $r_{ij}$ is assumed to be given by an RKKY interaction \cite{ruderman1954, fischer1975}, and the couplings extracted using DFT calculations provide information for fitting the function for general separations \cite{facemyer2023}. The inter-ion exchange takes the form
\begin{equation}\label{rkky}
    J(r_{ij}) = J_0 \frac{\cos (2k_F r_{ij} + \phi)}{r_{ij}^2} \,,
\end{equation}
where the Fermi-wavenumber for the substrate is $k_F = {2\pi}/{\lambda_F}$, and the Au(111) substrate has a surface-state Fermi-wavelength $\lambda_F = 36 \;\mathrm{\AA}$ \cite{sotthewes2021}. We find that $J_0 \simeq 2.752$ meV and $\phi \simeq 1.838$ provide an excellent fit to the DFT calculation \cite{facemyer2023}. For concreteness, we choose the vertex separations in the ferromagnetic and frustrated clusters as $b_{\mathrm{FM}}=5.01 \;\mathrm{\AA}$ and $b_{\mathrm{AFM}}=10.02 \;\mathrm{\AA}$, respectively, which correspond to inter-ion exchange interactions $J(b_{\mathrm{FM}}) \equiv J_1 = -0.101 \; \mathrm{meV}$ and $J(b_{\mathrm{AFM}}) \equiv J_2 = +0.0156 \; \mathrm{meV}$. We note that $b_{FM}$ corresponds to the separation between bridge sites on the Au(111) surface, and is close to the minimal inter-ion separation that can be resolved in STM experiments \cite{chen1999}.

Including the SIA in the Hamiltonian, as well as a Zeeman field, results in the system being described by
\begin{equation}\label{OurAHam}
    \hat{H}_0 = \sum_{i < j} J(r_{ij}) {\hat{\vec{S}}_{i} \cdot \hat{\vec{S}}_{j}} + A\sum_{i} \big(\hat{S}^{(z)}_i\big) ^2 - h_{\alpha}\sum_{i} \hat{S}^{(\alpha)}_i \, ,
\end{equation}
with `easy-plane' magnetic anisotropy $A = +0.05$ meV, as determined from DFT \cite{facemyer2023}; the Zeeman field $h_{\alpha}$ is applied in the $\alpha$-direction, with $\alpha \in \{x, z\}$. We use QuSpin \cite{weinberg2017} to diagonalize the Hamiltonian in full and evaluate the magnetic excitation spectrum; we are especially interested in the low-lying eigenenergies and eigenstates. Applying spin ladder and \textit{z}-projection operators to the ground state to mimic the action of tunneling electrons in STM experiments generates the tunneling spectral function.  This function describes the spin scattering via tunneling electrons when the STM tip is at ion $j$, so that
\begin{equation}\label{specfun}
\begin{split}
    A_j(E) =
    \sum_{\lambda}&\big\{ |\langle \lambda| \hat{S}_j^{+} |\lambda_0 \rangle|^2 + |\langle \lambda| \hat{S}_j^{-} |\lambda_0 \rangle|^2  \\ 
    &  +2 |\langle \lambda| \hat{S}_j^{z} |\lambda_0 \rangle|^2 \big\} \, \delta(E - E_{\lambda}),
    \end{split}
\end{equation}
where $\lambda$ denotes the eigenenergy basis of $H_0$ and $|\lambda_0\rangle$ represents the ground state \cite{ternes2015}. The tunneling spectral function $A_j(E)$ is proportional to the STM differential conductance at low bias voltages \cite{ternes2015,loth2010,schuh2012}.

The extent to which nonlocal correlations exist between subsystems is characterized using the von Neumann entanglement entropy \cite{srednicki1993},
\begin{equation}\label{vnEE}
     \mathcal{S}_{\mathrm{\gamma}} \equiv \mathcal{S}(\rho_{\mathrm{\gamma}}) = -k_B\mathrm{Tr}\big[ \rho_{\mathrm{\gamma}} \ln \rho_{\mathrm{\gamma}} \big] \, ,
\end{equation}
where $\rho_{\gamma}$ denotes the reduced density matrix of the subsystem under scrutiny and $k_B$ is the Boltzmann constant. We will report these values in terms of the maximal entropy of the system, $k_B \ln d_{\gamma}$, where $d_{\gamma}$ is the total number of quantum degrees of freedom (dimension) of subsystem $\gamma$. The reduced density matrix of the three-ion triangle subsystem A in Fig.\ \ref{system}, defined for the bipartite system $A\oplus B$ as $\rho_{A} \equiv \mathrm{Tr}_B (\rho_{AB})$, with $\rho_{AB} \equiv |\psi_{AB}\rangle \langle \psi_{AB}|$, has $d_{A} = (2S+1)^3= 8^3 $ for $S=7/2$. In a pure system, $\mathcal{S} = 0$ and $\mathrm{Tr}(\rho^2) = 1$. We are interested in establishing the entanglement of subsystems of the pure bipartite system $\rho_{AB}$, so that $\mathcal{S}_{A} > 0$ and $\mathrm{Tr}(\rho_{A}^2) < 1$ for different configurations of subsystem A and probe-ion $\{3\}$, as in Fig.\ \ref{system}.

%%%%%%%%%%%%%%%%%%%%%%%%%%%%%%%%%%%%%%%%%%%%
\section{\label{Results}Magnetic excitations and spectroscopy}

%%%%%%%%%%%%%%%%%%%%%%%%%%%%%%%%%%%%%%%%%%%%
\subsection{Magnetic excitation spectrum}
\begin{figure}[h]
    \begin{overpic}[width=1\columnwidth]{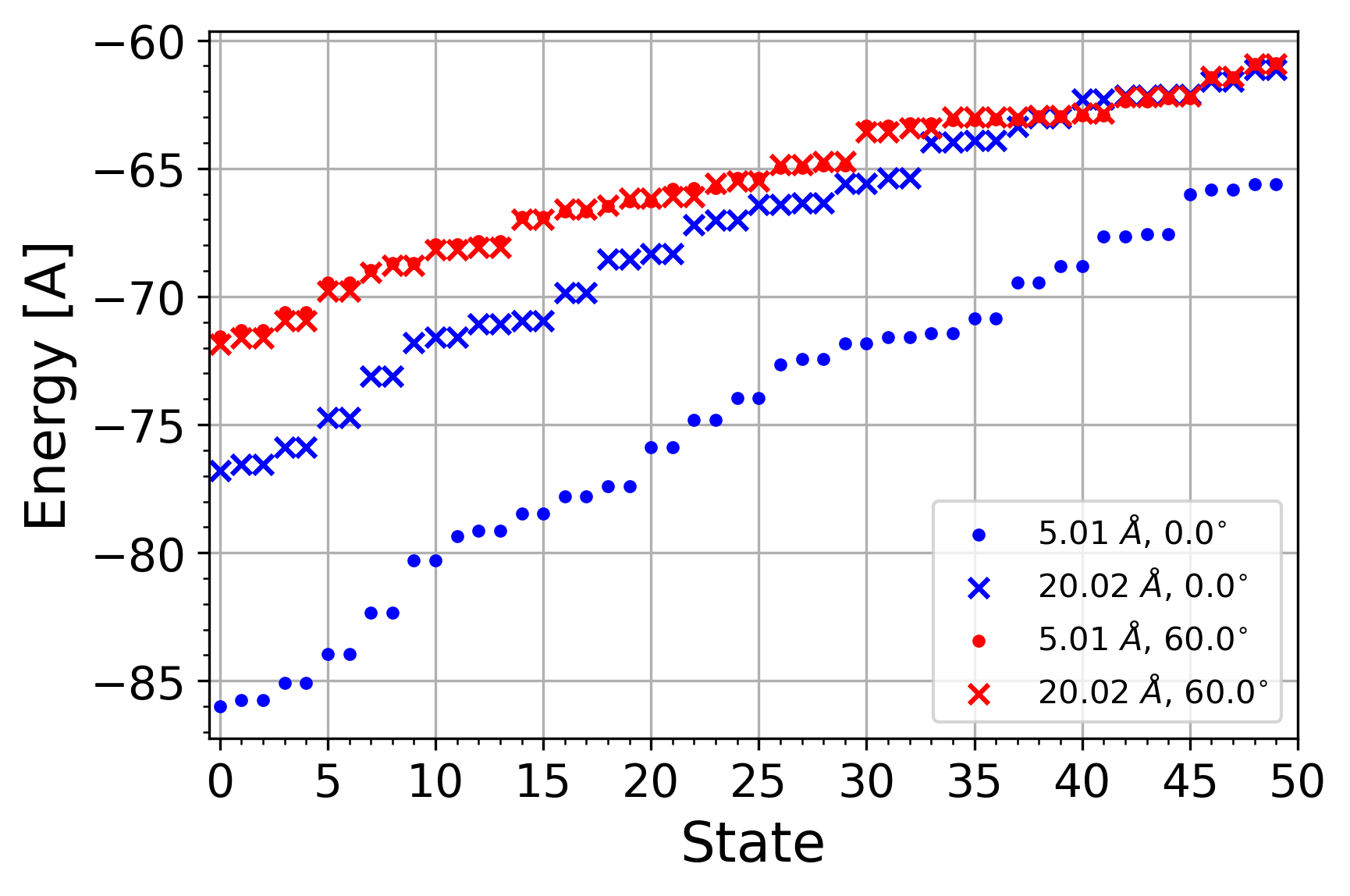}%
    %\put(520,120){\frame{\includegraphics[width=4.1cm]{fm_en_spec_per_pos.png}}}%
    %\put(10,620){(1)}%
    %\put(560,360){(2)}%
    \end{overpic}
    \caption{Ferromagnetic cluster. Low-lying magnetic excitation spectrum for the four-atom cluster of $S = 7/2$ magnetic moments. The triangular cluster (subsystem A) is composed of vertex ions separated by $b_{FM}\approx 5 \,\mathrm{\AA}$, coupled ferromagnetically with $J_1 = -0.101 \; \mathrm{meV}$. Ground states for different probe-ion (subsystem B) positions are non-degenerate due to the SIA contribution, $A=0.05$ meV. Energies in units of $A$. Legend indicates ($r',\theta$) values for the probe-ion in different configurations. \label{FMSpectrum}}
\end{figure}

In the first configuration case we consider, the relatively large ferromagnetic coupling $J_1 = -0.101 \; \mathrm{meV}$ is associated with bridge-site separation of $b_{FM}\approx 5 \;\mathrm{\AA}$ between ions in the triangular cluster and dominates the system. This strong \textit{ferromagnetic} coupling in the triangular cluster results in the low-lying excitation spectra for various probe-ion positions shown in Fig.\ \ref{FMSpectrum}. Ground state manifolds with maximal spin, that would otherwise be highly degenerate without SIA, are split due to the comparable strength of the easy-plane SIA. The strong SIA results in the ground state being a ferromagnetic singlet with $S_z=0$, while the total spin $S_{\mathrm{total}}=14$. The next higher doublet has $S_z=\pm 1$, $S_{\mathrm{total}}=14$, etc. The probe-ion influence on subsystem A enhances the SIA splitting by further splitting energy manifolds. When the SIA is set to zero, the probe-ion exchange is not strong enough in any of the four arrangements shown to split the ground state manifold; however, the probe-ion is responsible for clear shifts of the spectra for different probe-ion positions. The exchange couplings acting on the probe-ion play a significant role in the case when the probe is at $(r'\approx 5\,\mathrm{\AA}, \theta \approx 0^{\circ})$. The other probe-ion positions considered in Fig.\ \ref{FMSpectrum} show a gradual clustering of the energy levels, indicating that the probe-ion becomes only mildly perturbative at larger distances.

\begin{figure}[h]
    \begin{overpic}[width=1\columnwidth]{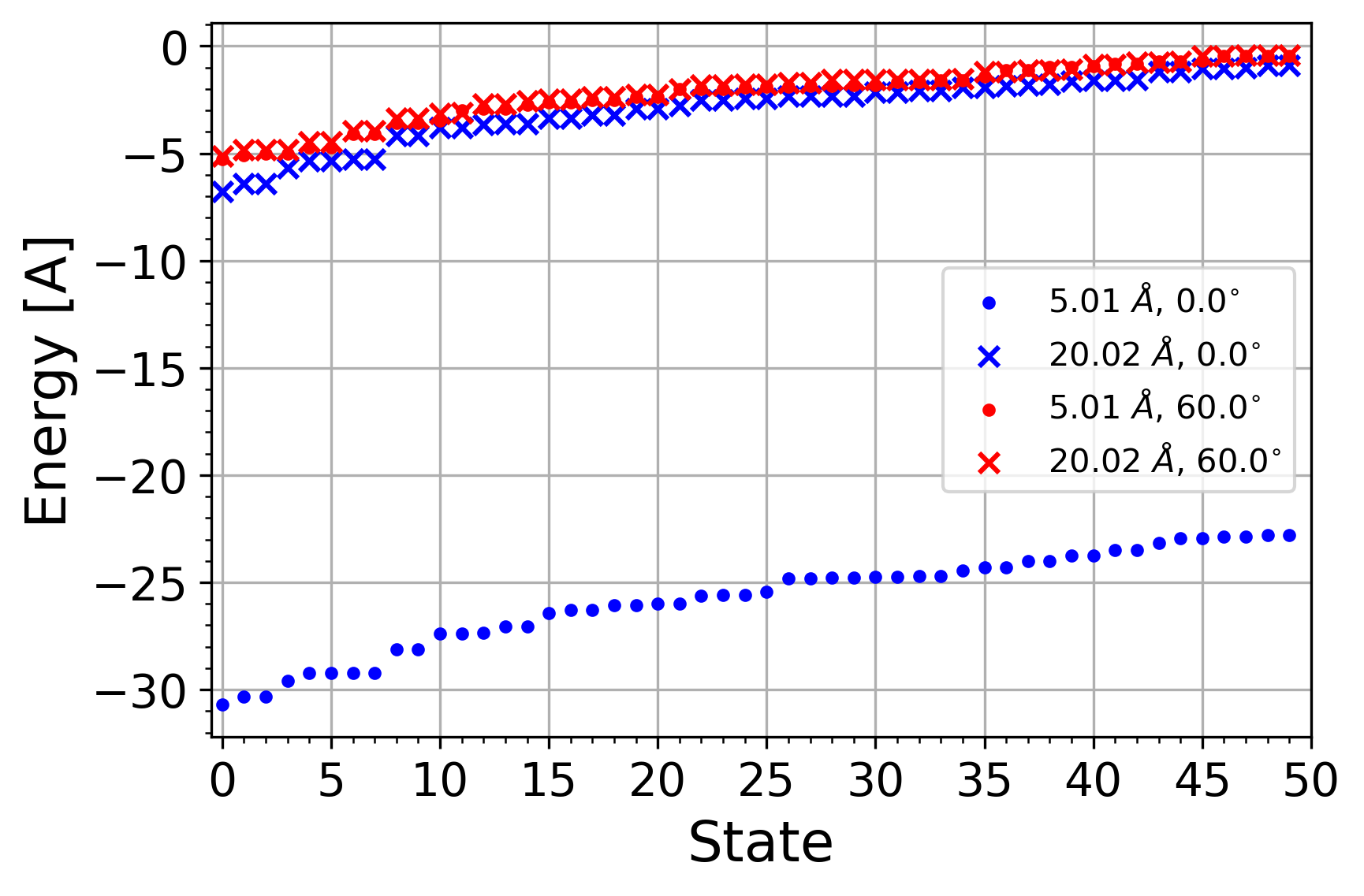}%
    %\put(520,120){\frame{\includegraphics[width=4.1cm]{fm_en_spec_per_pos.png}}}%
    %\put(10,620){(1)}%
    %\put(560,360){(2)}%
    \end{overpic}
    \caption{Antiferromagnetic cluster. Low-lying magnetic excitation spectra for the four-adatom cluster of $S = 7/2$ magnetic moments. The triangular cluster (subsystem A) is composed of vertex ions separated by $b_{AFM} \approx 10 \,\mathrm{\AA}$, coupled antiferromagnetically with $J_2 = +0.0156 \; \mathrm{meV}$. Ground states for different probe-ion (subsystem B) positions are non-degenerate due to the SIA contribution, $A=0.05$ meV. Energies in units of $A$. Legend indicates ($r',\theta$) positions of the probe-ion. \label{AFMSpectrum}}
\end{figure}

The second case considered is a triangular cluster that has antiferromagnetically-coupled vertex ions separated by $b_{AFM} \approx 10 \,\mathrm{\AA}$, with strength $J_2 = +0.0156 \; \mathrm{meV}$. The corresponding low-lying excitation spectra for various probe-ion positions are shown in Fig.\ \ref{AFMSpectrum}. The ground state is here also a singlet, on account of the dominant SIA present. The spectra are clustered more overall than before, as the smaller $J_2$-values result in weaker splittings. Similarly, the energy binding is smaller. The probe-ion at $(r',\theta) \approx (5 \,\mathrm{\AA}, 0^{\circ})$ is $J_1$-coupled (ferromagnetically) with the \{0\} vertex-ion, which yields the energetically favored geometry seen in Fig.\ \ref{AFMSpectrum}.
Other probe-ion positions appear to be only weakly perturbative, especially for $\theta = 60^\circ$.

%%%%%%%%%%%%%%%%%%%%%%%%%%%%%%%%%%%%%%%%%%%%
\subsection{Tunneling spectroscopy and applied field}
As we will show, the differential conductance profiles and entanglement entropy (EE), when plotted as functions of applied field, show clear and striking connections.

\begin{figure}[h]
    \begin{center}
        \begin{overpic}[width=0.52\columnwidth]{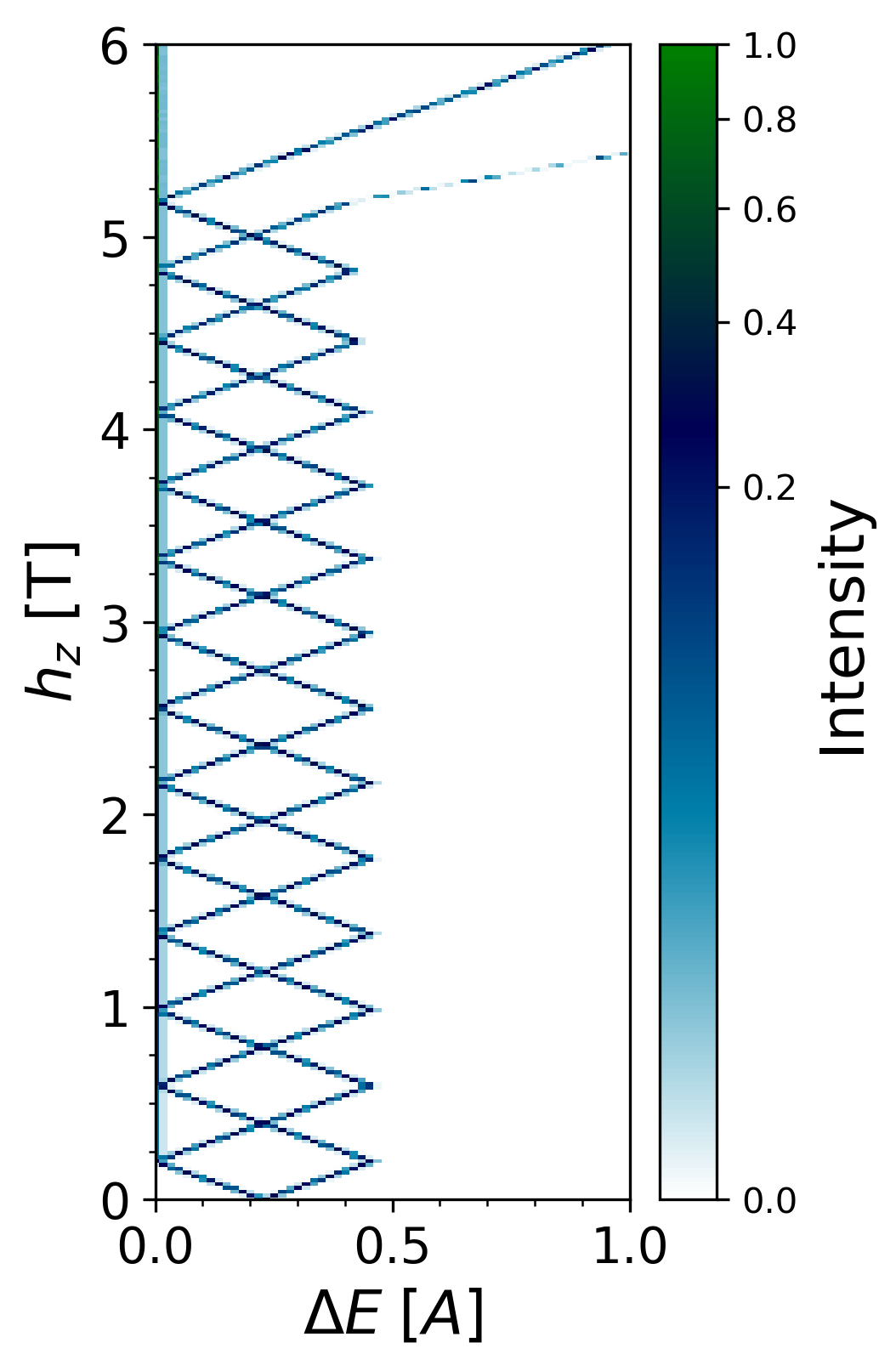}%
        \put(150,900){(a)}%
        \end{overpic}
        \begin{overpic}[width=0.45\columnwidth]{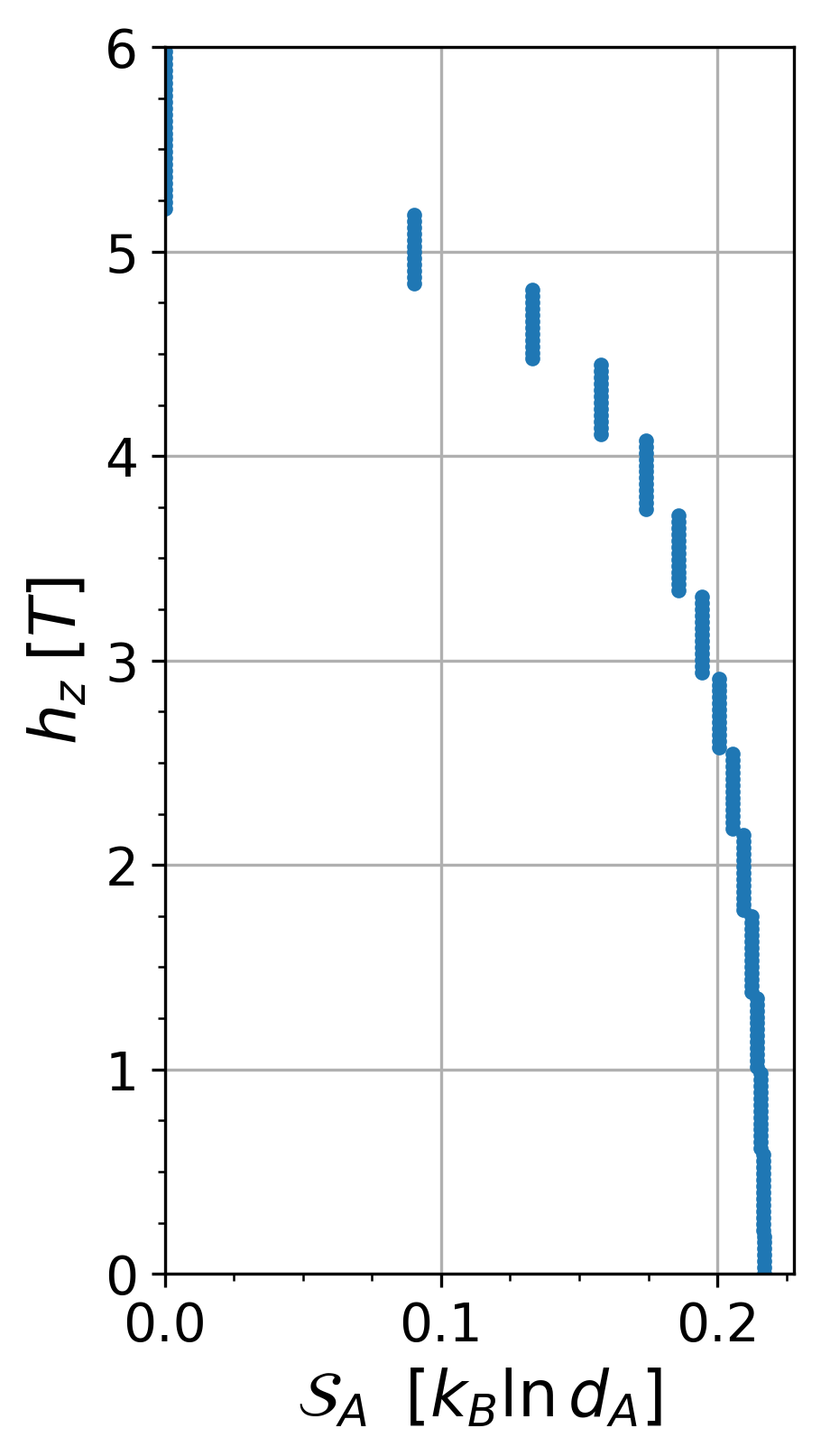}%
        \put(150,900){(b)}%
        \end{overpic}
        \begin{overpic}[width=0.52\columnwidth]{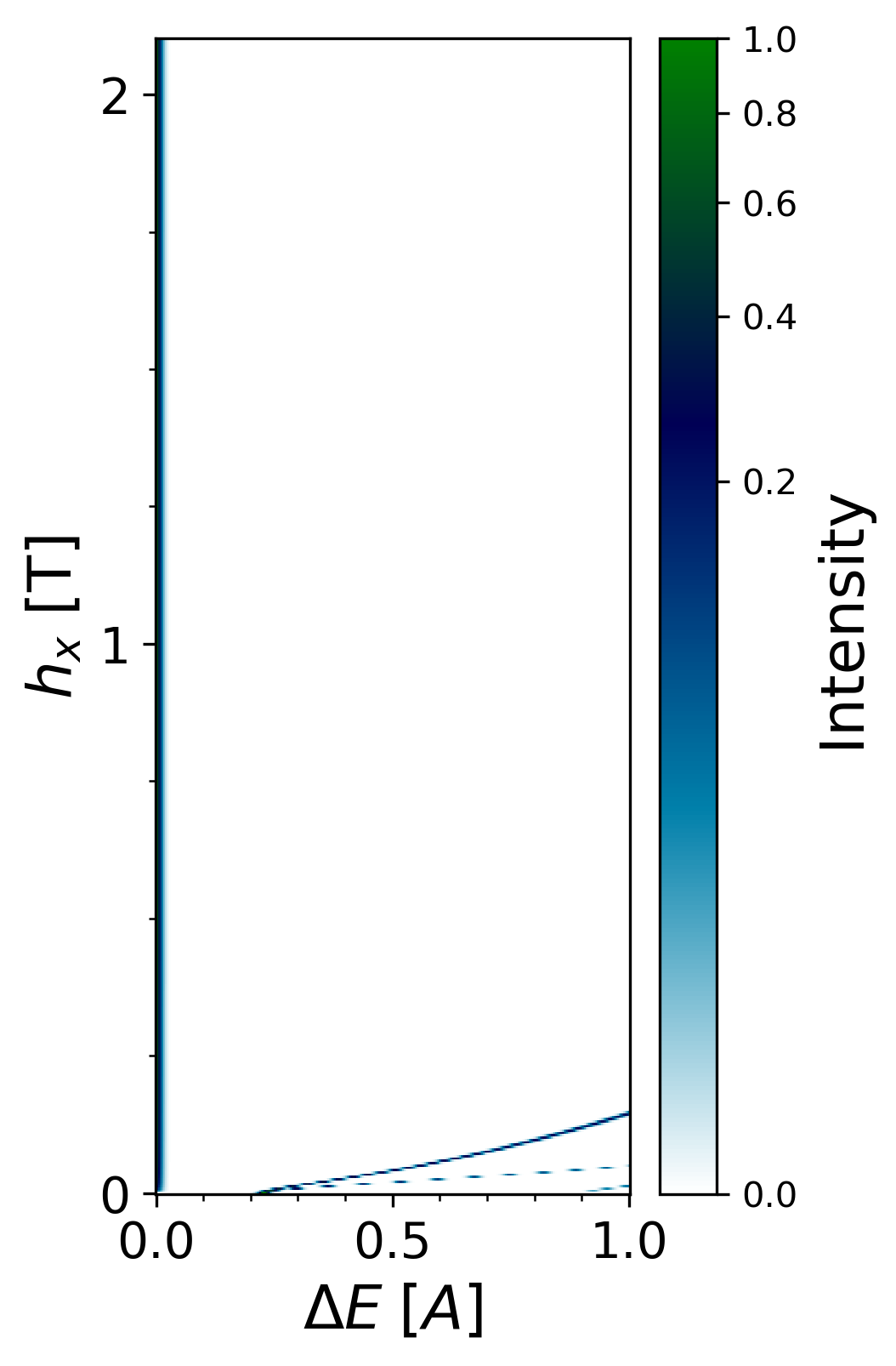}%
        \put(150,900){(c)}%
        \end{overpic}
        \begin{overpic}[width=0.45\columnwidth]{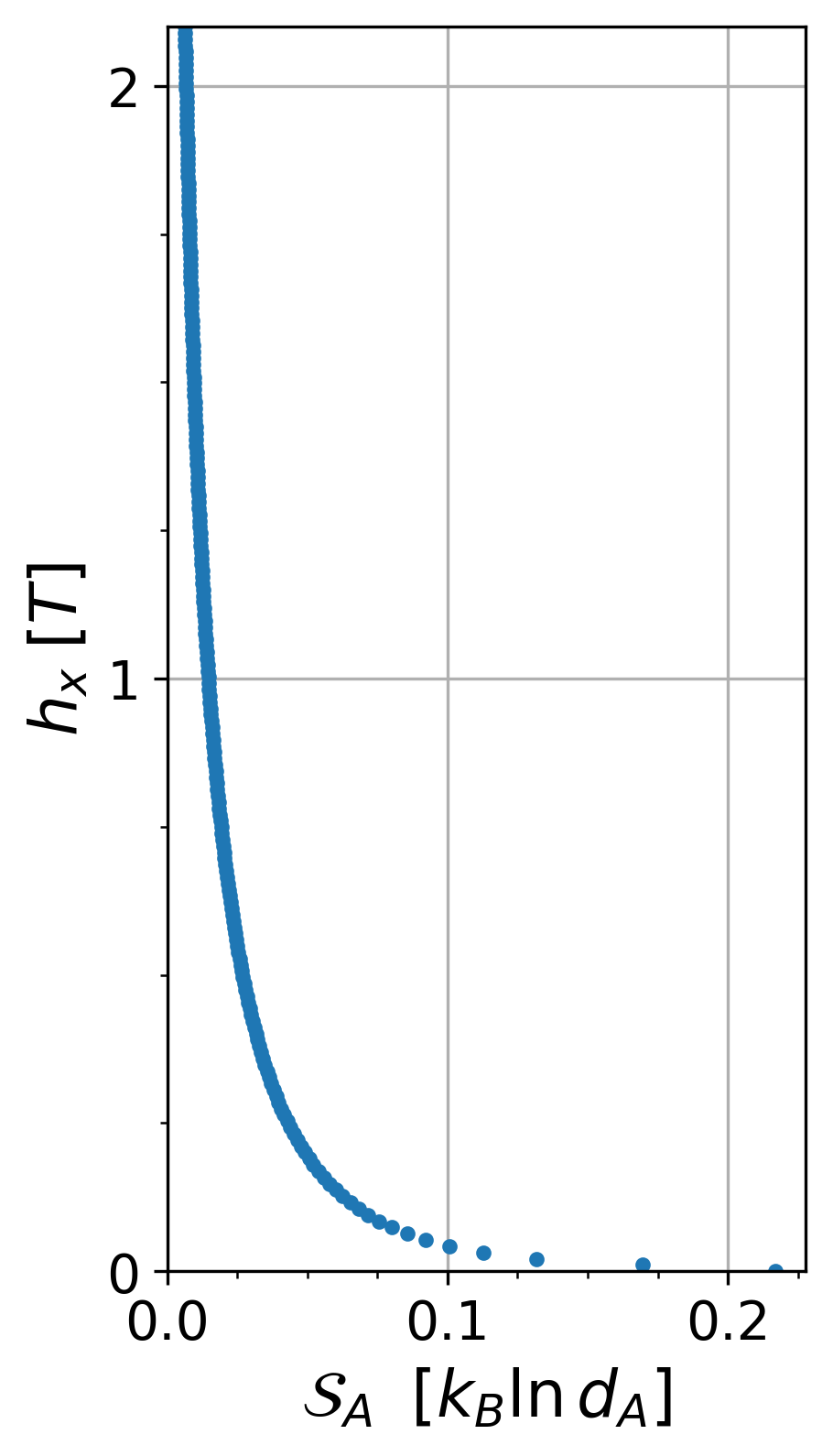}%
        \put(150,900){(d)}%
        \end{overpic}
    \end{center}
\caption{Ferromagnetic cluster. (a,c) color maps of the tunneling spectral function (proportional to differential conductance) for the ferromagnetic triangular cluster and Zeeman field along different directions. The probe-ion is strongly coupled ferromagnetically to the zeroth ion ($r'\approx 5$ \AA, $\theta = 0^{\circ}$). The tunneling is on the zeroth ion. Corresponding entanglement entropies are shown in (b,d), where the probe-ion has been traced out.} \label{FM_dcee_field}
\end{figure}

We return to the first case, where subsystem A constitutes the strongly coupled ferromagnetic triangular cluster and consider the probe-ion is placed $b_{FM}$ away from the 0-vertex ion, so that they couple ferromagnetically with the same strength $J_1$. We see from Fig.\ \ref{FM_dcee_field}(a,b) that an external out-of-plane magnetic field ($\alpha=z$) produces a \textit{braiding} structure in the differential conductance profile (i.e. peak crossings over extended field range and energy) that directly correspond to non-zero values of EE. The braiding is produced by the increasing polarization of the system as the field increases, so that the $S_z$ eigenvalue of the ground state goes successively from 0 to 1, 2, etc. These $S_z$ transitions extend to a field-depth of $h_z \simeq 5$ T, when the large field overtakes the SIA and the various exchange couplings, producing a fully polarized, fully factorizable ferromagnetic ground state with $S_z=S_{\mathrm{total}}=14$. Plateaus in the EE appear in field windows over which the ground state has a definite $S_z$, with low-energy transitions changing linearly in field, before another step increases the polarization and reduces the EE accordingly.

To guide our intuition concerning braiding and its connection to nonlocal correlations, it helps to remember what peaks in the differential conductance profiles say about the corresponding state. Peaks in $A_j(E)$ indicate that the STM-tip electrons produce an excited state ($\Delta E>0$) with probability proportional to the peak height. Recall that the flip-state reached after the tip electron scatters from the magnetic ion is not an eigenstate of the system in general, but a linear combination of the excited eigenstate manifold. If the ground state is a non-separable superposition of spin states, especially as in the case where the ground state has small $S_z$, we would expect low-lying excited state peaks to follow linear Zeeman splitting trajectories when external fields are present. As the field increases and polarizes the system, the $S_z$ of the ground state increases in steps. For characteristic field depths, one would see repeated energy-level crossings, producing the braiding we see in $A_j(E)$. If the ground state is separable, like in the case of a fully polarized ground state, one will {\em not} see braiding because the energy gaps between the ground state and accessed eigenstates after single spin flips increase proportionally to field strength. In other words, there will not be crossings of the $A_j(E)$ peaks. 

When the Zeeman field is applied in-plane ($\alpha=x$) to the same system, a different behavior emerges, as seen in Fig.\ \ref{FM_dcee_field}(c,d). The EE is quickly and continuously reduced (as opposed to the discrete steps in the previous case) with growing field strength. As the ground state has $S_{\mathrm{total}}=14$, $S_z=0$, the $S_x$ polarization of the ground state is gradually enhanced by the in-plane field, which suppresses the EE. The dramatically different structure compared to the out-of-plane field highlights the role that the easy-plane SIA plays and provides an alternative way to probe the nature of magnetic anisotropy in a given system.

\begin{figure}[h]
    \begin{center}
        \begin{overpic}[width=0.52\columnwidth]{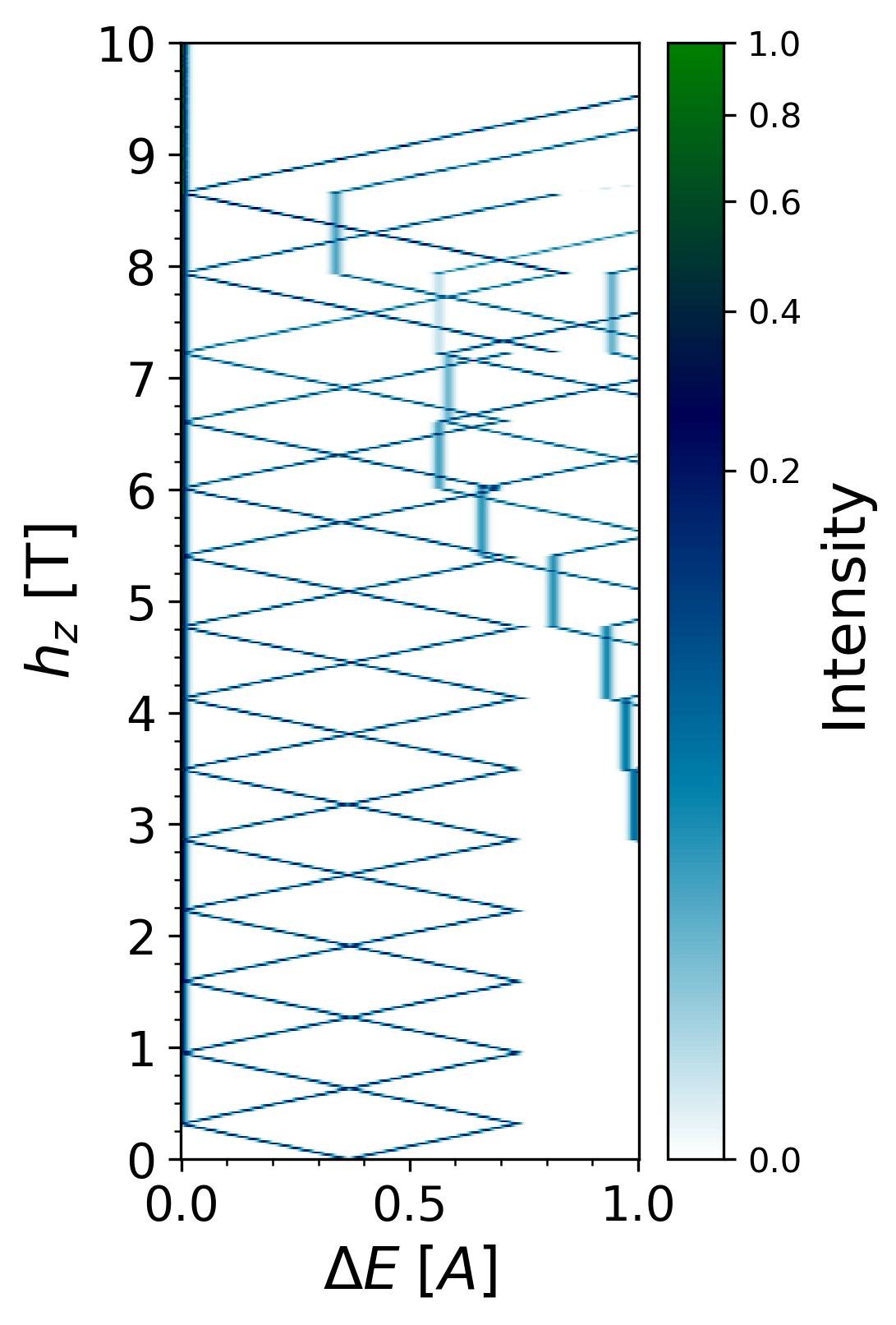}%
        \put(150,900){(a)}%
        \end{overpic}
        \begin{overpic}[width=0.45\columnwidth]{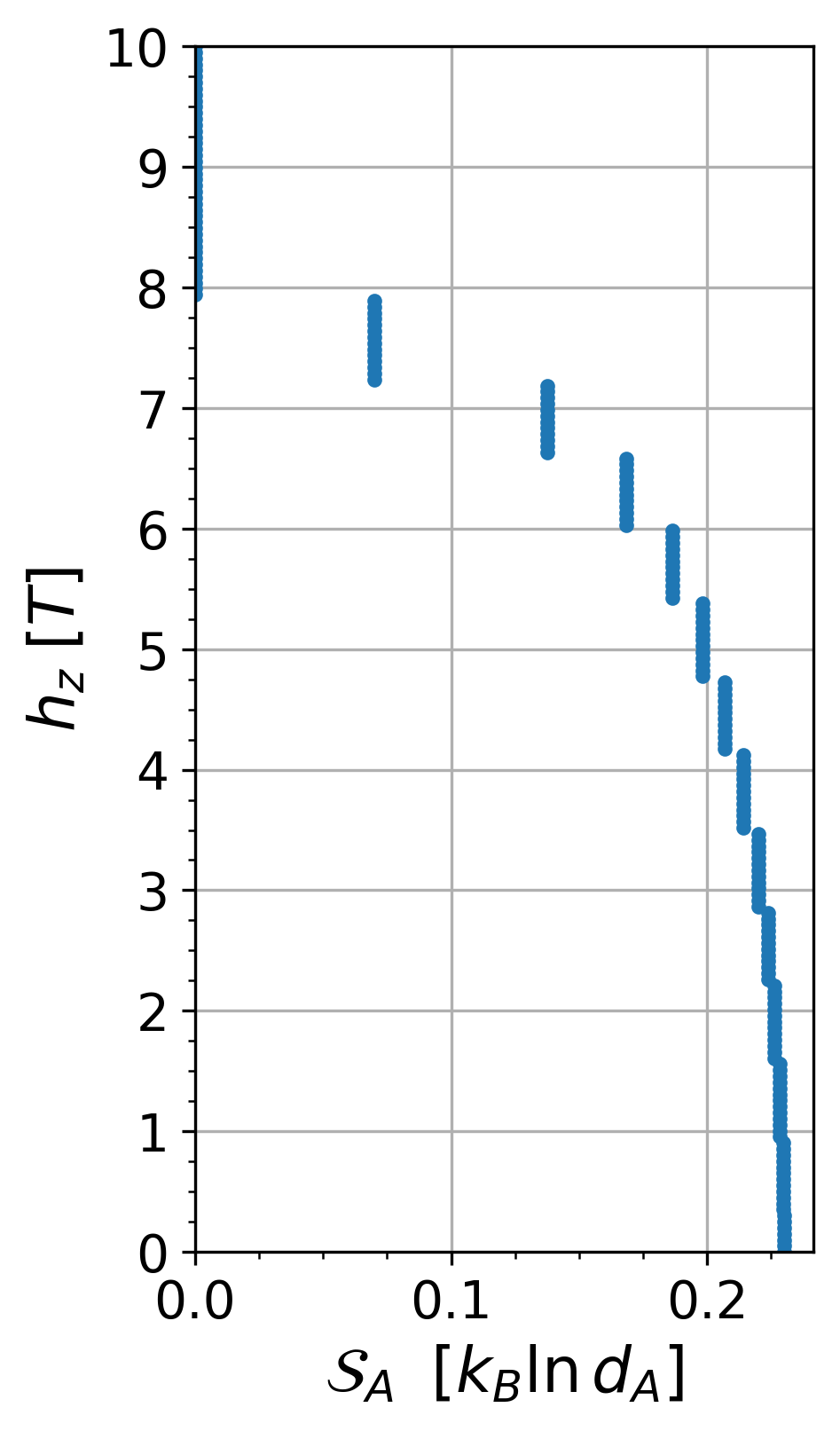}%
        \put(150,900){(b)}%
        \end{overpic}
        \begin{overpic}[width=0.52\columnwidth]{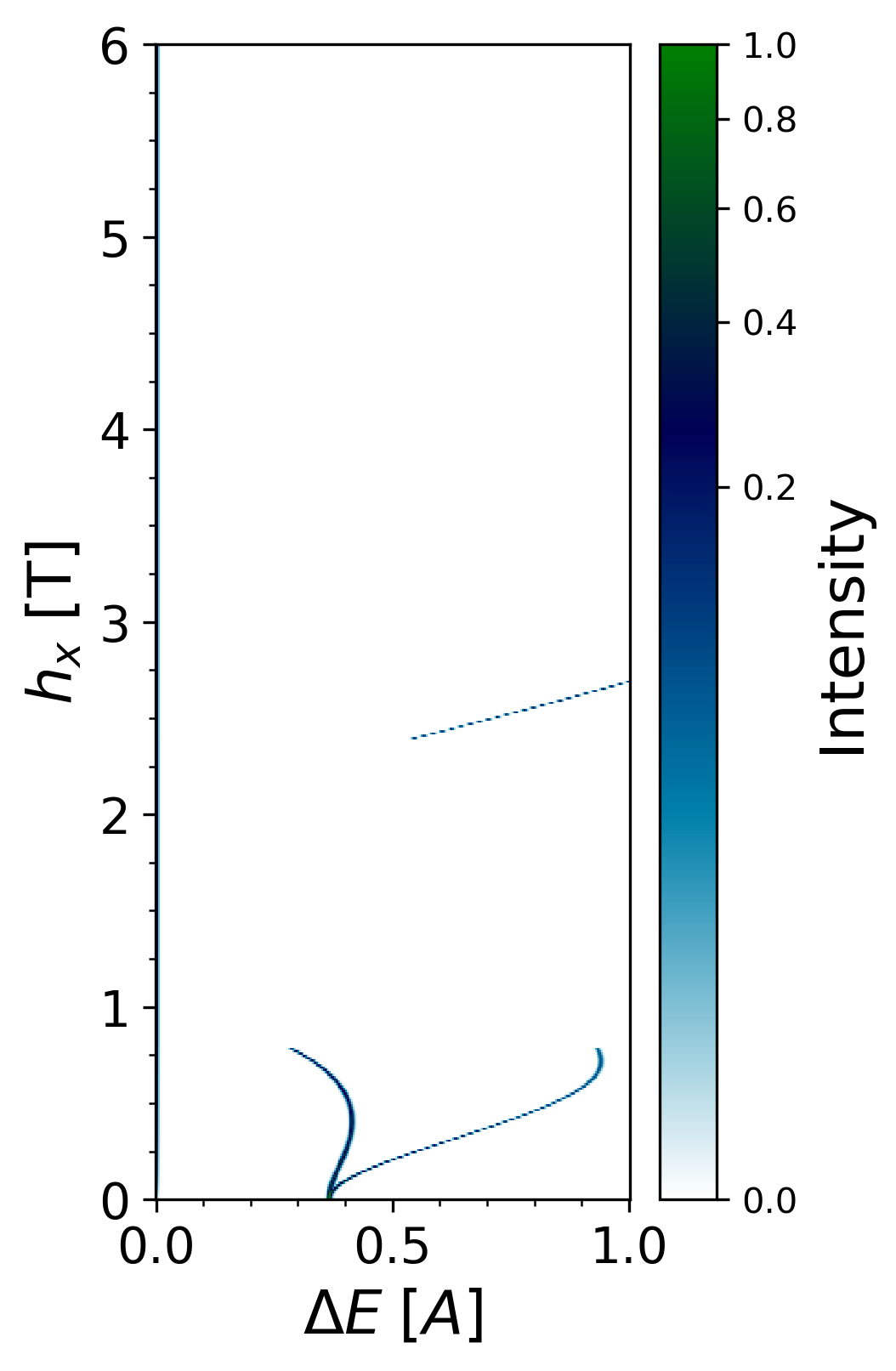}%
        \put(150,900){(c)}%
        \end{overpic}
        \begin{overpic}[width=0.45\columnwidth]{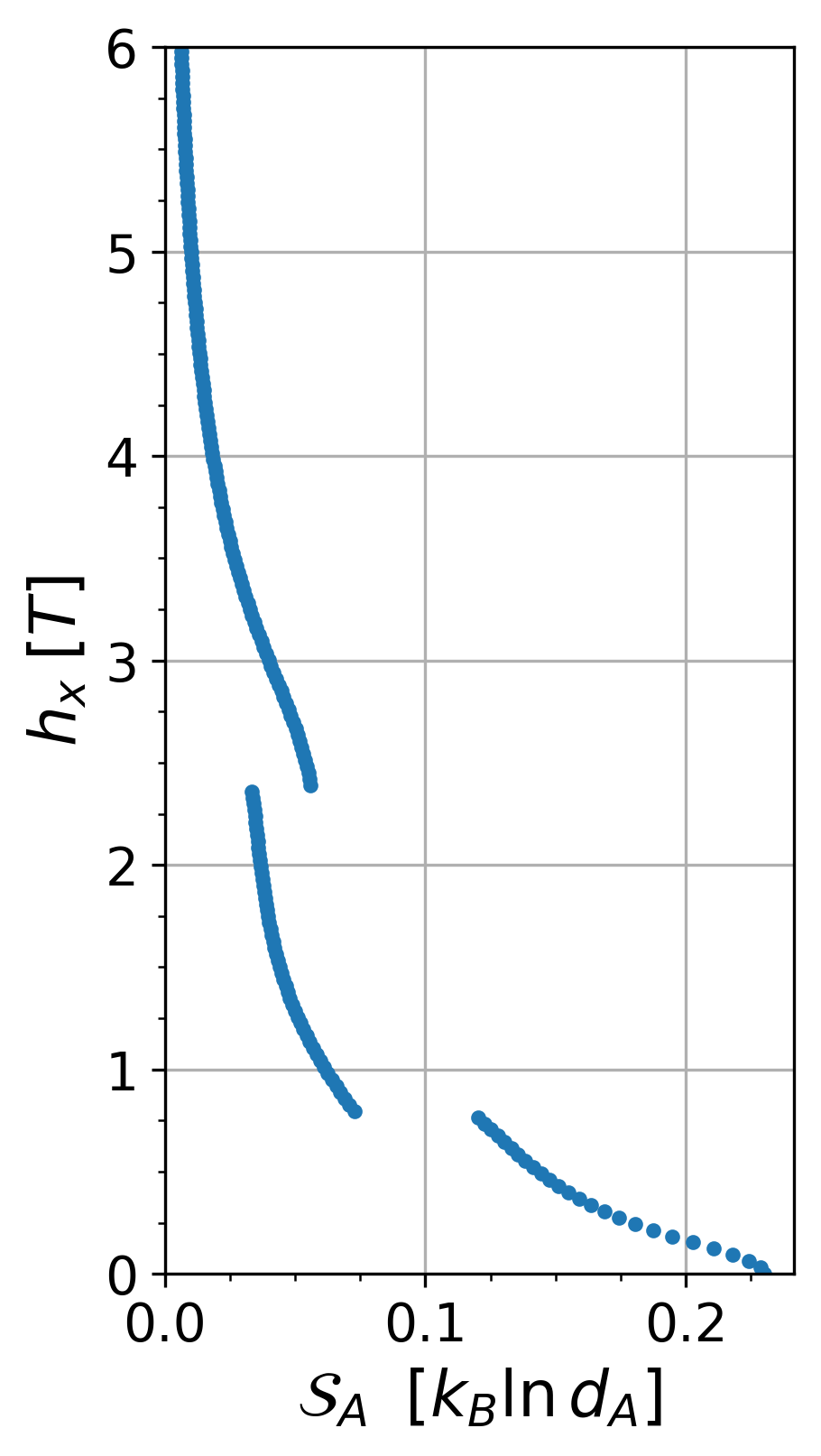}%
        \put(150,900){(d)}%
        \end{overpic}
    \end{center}
\caption{Antiferromagnetic cluster. (a,c) color maps of the tunneling spectral function (proportional to differential conductance) for the antiferromagnetic triangular cluster and Zeeman field along different directions. The probe-ion is coupled ferromagnetically to the zeroth ion ($r'\approx 5$ \AA, $\theta = 0^{\circ}$). The tunneling is on the zeroth ion. Corresponding entanglement entropies are plotted alongside (b,d), where the probe-ion $\{3\}$ has been traced out.} \label{AFM_dcee_field}
\end{figure}

The second configuration, where subsystem A is made up of a frustrated triangular cluster with antiferromagnetic couplings $J_2$, while the probe-ion is ferromagnetically coupled to the 0-vertex ion with $J_1$, shows similar braiding behavior with $h_z$-field, as seen in Fig.\ \ref{AFM_dcee_field}. The overall characteristics are quite similar to what we saw in the ferromagnetic cluster, although one notices some differences in the entanglement measure for stronger fields. The \textit{braiding} features in the differential conductance characterize a field-depth over which EE for subsystem A is nonzero, decreasing in discrete steps as the differential conductance indicates increasing polarization of the ground state. It is interesting to note that the EE vanishes earlier in Fig.\ \ref{AFM_dcee_field}(b) than the braiding in (a) would suggest. This is an example when braiding in the differential conductance informs us that there is a part of the system that still has nonzero EE, even as the probe-ion is traced out. In this case, the braiding in \ref{AFM_dcee_field}(a) indicates the EE present {\em within} subsystem A terminates at about $8.7$ T. A one-tangle calculation, where subsystem A is ion $\{2\}$ in the cluster (i.e., tracing out \{0,1,3\}), has EE that in fact terminates at the field strength suggested by the braiding, $\approx 8.7$ T, when the triangular cluster is fully polarized.

It is important to emphasize that the connection between EE and braiding in $A_j(E)$ seen for increasing $h_z$ field requires the purity of the ground state we assume in Fig.\ \ref{FM_dcee_field} and \ref{AFM_dcee_field}.  In contrast, a fully excited state, thermalized at high temperatures, such that all eigenstate transitions are possible, has no correlations and no EE.  Yet, it exhibits a multitude of braiding patterns in $A_j(E)$. The required low-temperature conditions to ensure a pure ground state are possible in the typical environments of high-precision STM experiments.

%%%%%%%%%%%%%%%%%%%%%%%%%%%%%%%%%%%%%%%%%%%%
\subsection{Spectroscopy and geometry}
So far, we have shown that a braiding structure in the differential conductance profiles in the presence of applied magnetic fields orthogonal to the SIA is accompanied by changes in the EE of the chosen subsystem A. One can clearly point to regimes of nontrivial entanglement within the bipartite system as the differential conductance changes. This is remarkable in its own right, but a quantitative measure of some sort would be desirable to complement these results. We now describe how to obtain such measures when we survey the differential conductance not with applied magnetic field but with probe-ion separation.

Figure \ref{ee_geo}(a) is a contour map of the EE for the ferromagnetic cluster of Fig.\ \ref{FM_dcee_field}, while Fig.\ \ref{ee_geo}(c) shows EE for the antiferromagnetic cluster of Fig.\ \ref{AFM_dcee_field}. In both maps we allow the probe-ion, i.e., subsystem B, to canvass the region around the cluster for $(r',\theta)$ positions that span the region illustrated in Fig.\ \ref{system}. The EE in either scenario exhibits oscillations as a function of both distance $r'$ and angle $\theta$. Although the SIA safeguards the entanglement between subsystems regardless of the probe-ion position (as it is not affected by the separation between subsystems) the decreasing RKKY couplings present produce oscillations with the characteristic period of $\lambda_{F}/2= 18$ \AA\ \cite{sotthewes2021}. To illustrate the connection between the EE variations and RKKY couplings, Fig.\ \ref{ee_geo}(b) and \ref{ee_geo}(d) show corresponding contour plots of a measure of the exchange coupling between subsystems A and B, which we obtain by adding the exchanges between each ion in subsystem A and the probe-ion (i.e., subsystem B), $J_3 = \sum_l J_{l3}=J_{03} + J_{13} + J_{23} $. Although a simple estimate, it captures the main features of the modulated behavior of the EE in both cases. {We see that successive maxima in EE (light blue regions in \ref{ee_geo}(a) and (c)) are due to extrema (max and min) in the effective exchange (positive and negative exchange values correspond to purple and green regions, respectively, with  white corresponding to zero $J_3$ value). Minima in the EE occur whenever the effective exchange is near zero and the probe is nearly decoupled.

If we examine the system without SIA, there are in fact probe-ion separations (as $J_3 \approx0)$ where the effective subsystem couplings produce separable states, i.e., zero EE. On the other hand, as discussed above, a nonzero easy-plane SIA energetically favors the lowest $S_z$ of a given $S_{\mathrm{total}}$ state, ensuring that a nonzero EE remains in the ground state, as seen in Fig.\ \ref{ee_geo}. An asymptotic non-vanishing bound in EE is the signature for topological EE discussed in the literature \cite{kitaev2006}, and this RE ion cluster is an interesting microscopic model on which to study its properties.
\begin{figure}[h]
    \begin{center}
        \begin{overpic}[width=0.48\columnwidth]{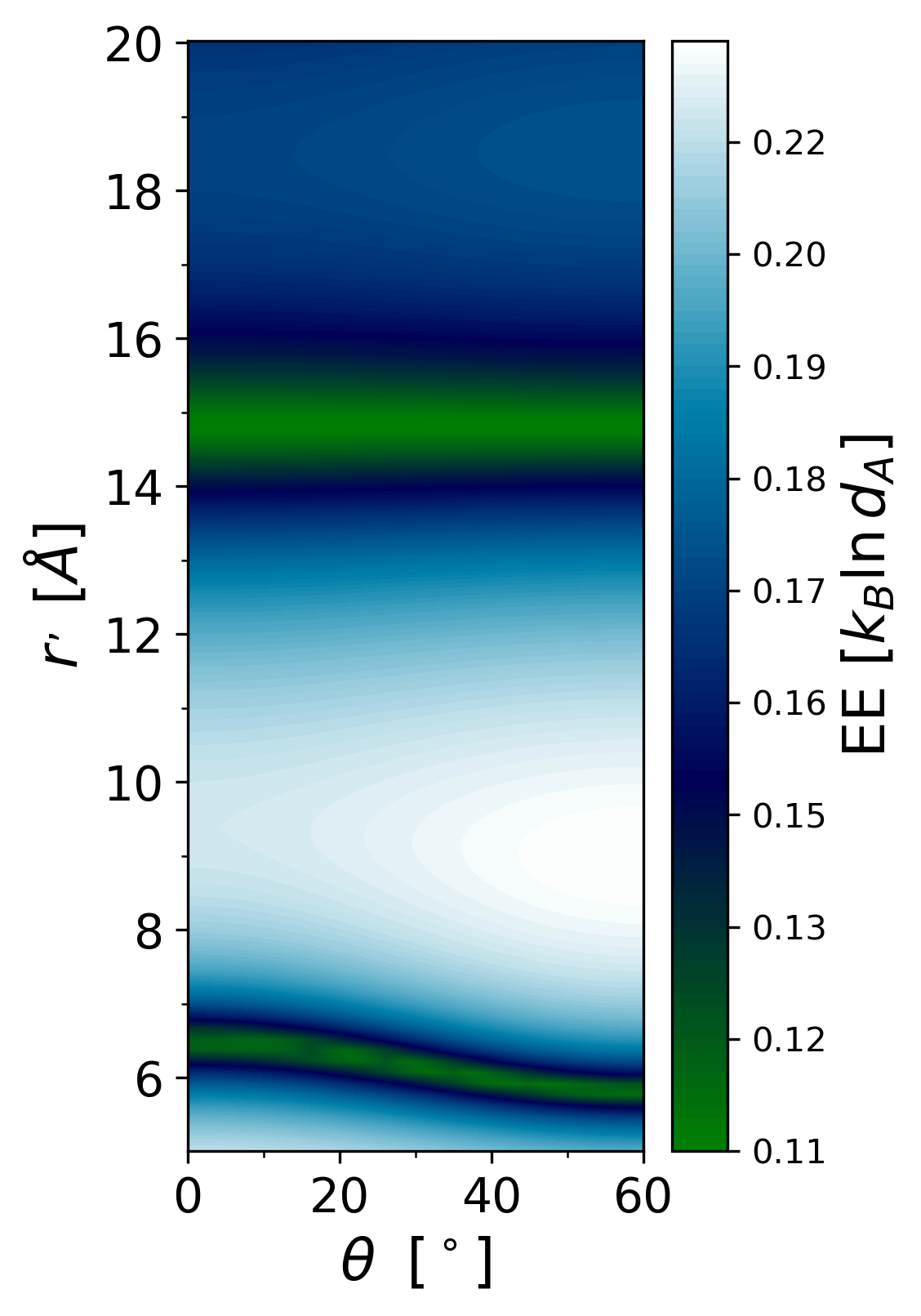}%
        \put(150,900){\textcolor{white}{(a) FM}}%
        \end{overpic}
        \begin{overpic}[width=0.49\columnwidth]{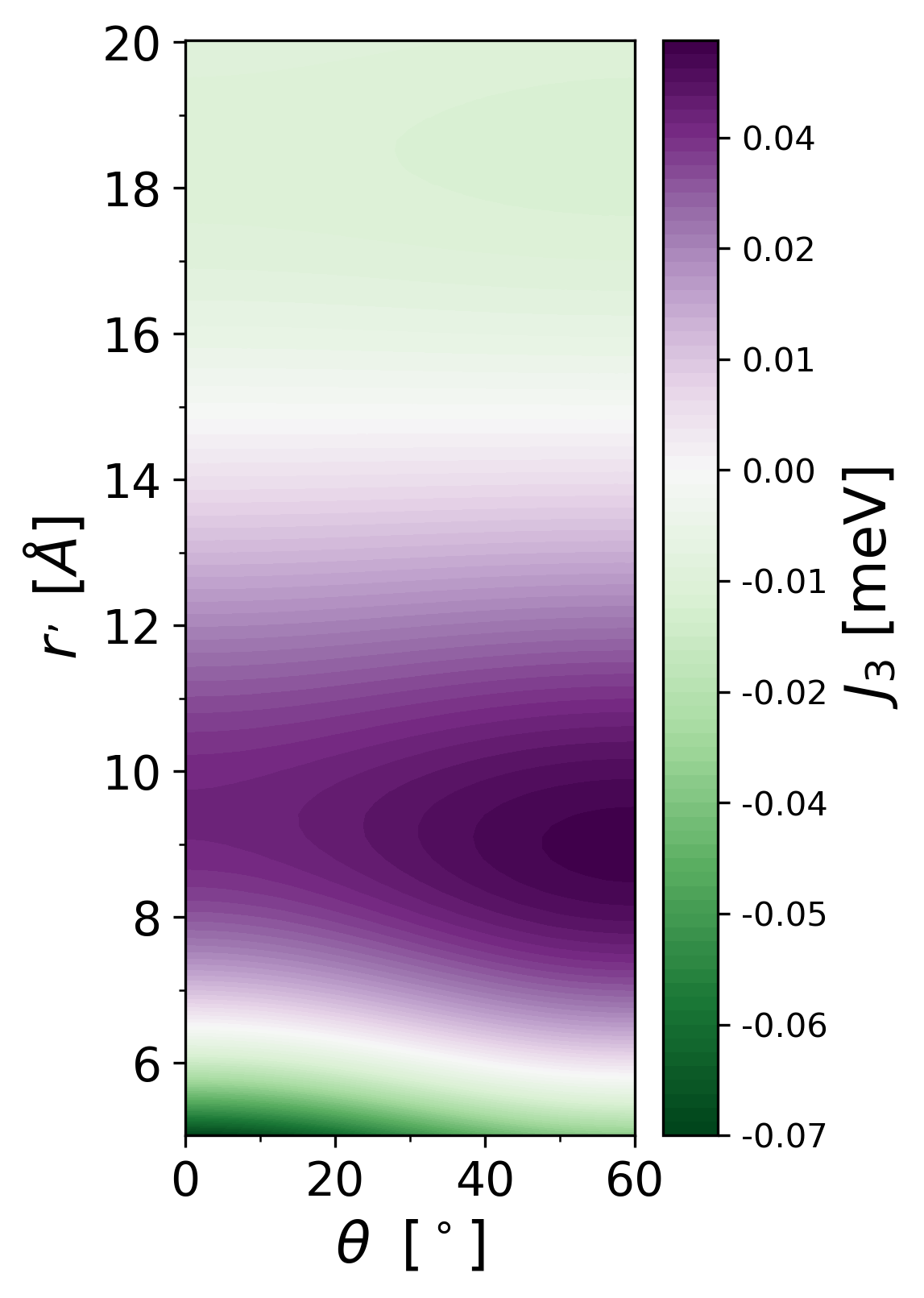}%
        \put(150,900){(b) FM}%
        \end{overpic}
        \begin{overpic}[width=0.48\columnwidth]{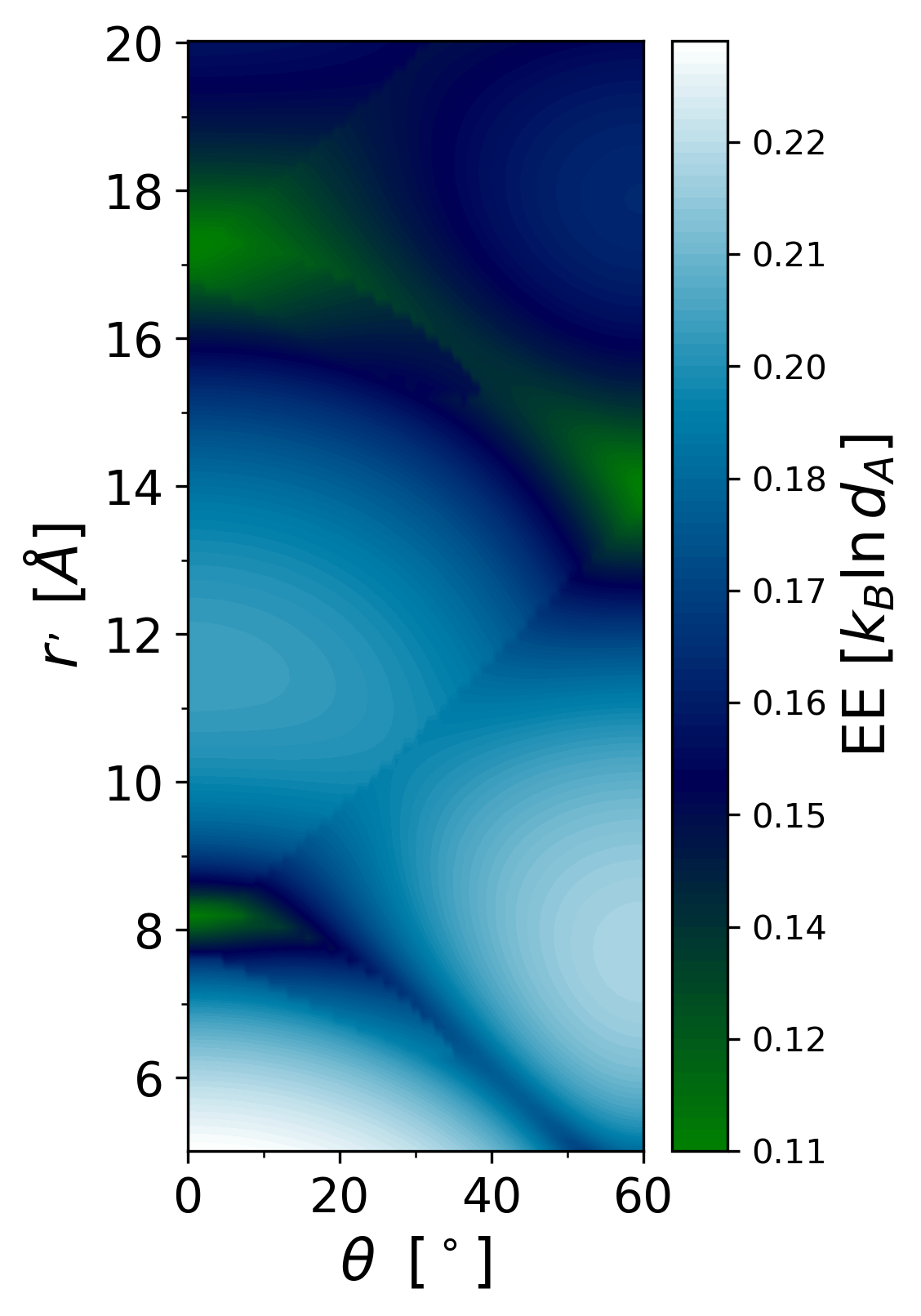}%
        \put(150,900){\textcolor{white}{(c) AFM}}%
        \end{overpic}
        \begin{overpic}[width=0.49\columnwidth]{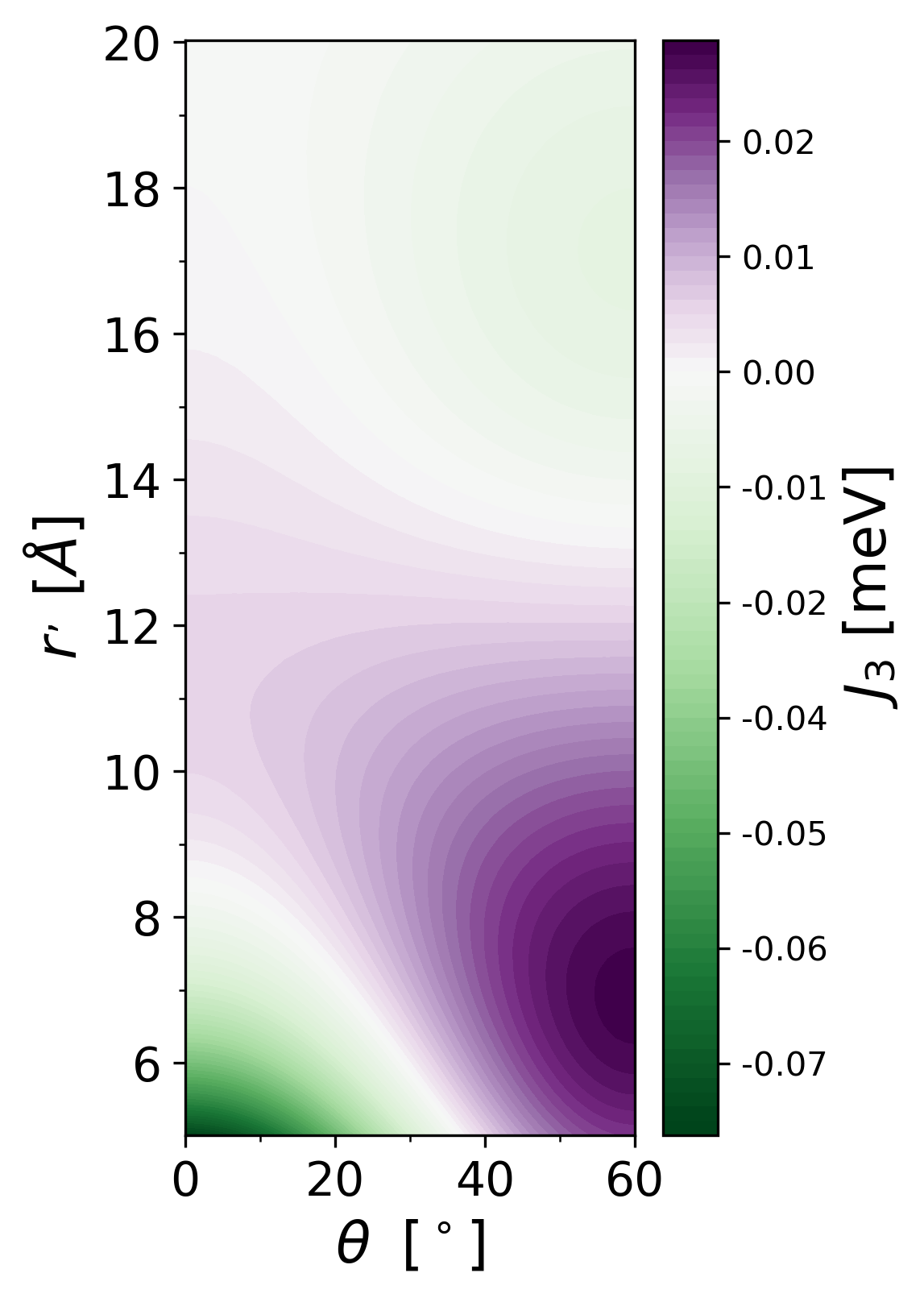}%
        \put(150,900){(d) AFM}%
        \end{overpic}
    \end{center}
\caption{EE color maps for both (a) the ferromagnetically (FM) coupled and (c) the antiferromagnetically (AFM) coupled  triangular clusters (subsystem A) as functions of the probe-ion (ion \{3\}) position ($r',\theta$) away from the \{0\}-ion in the cluster, as indicated in Fig.\ \ref{system}. Corresponding \textit{effective} exchange couplings plots $J_3$ (b,d) as defined in the text. Variations in $J_3$ clearly correlate with those in EE, as minima in EE occur whenever $J_3$ vanishes.} \label{ee_geo}
\end{figure}

\begin{figure}[h]
    \begin{center}
        \begin{overpic}[width=0.52\columnwidth]{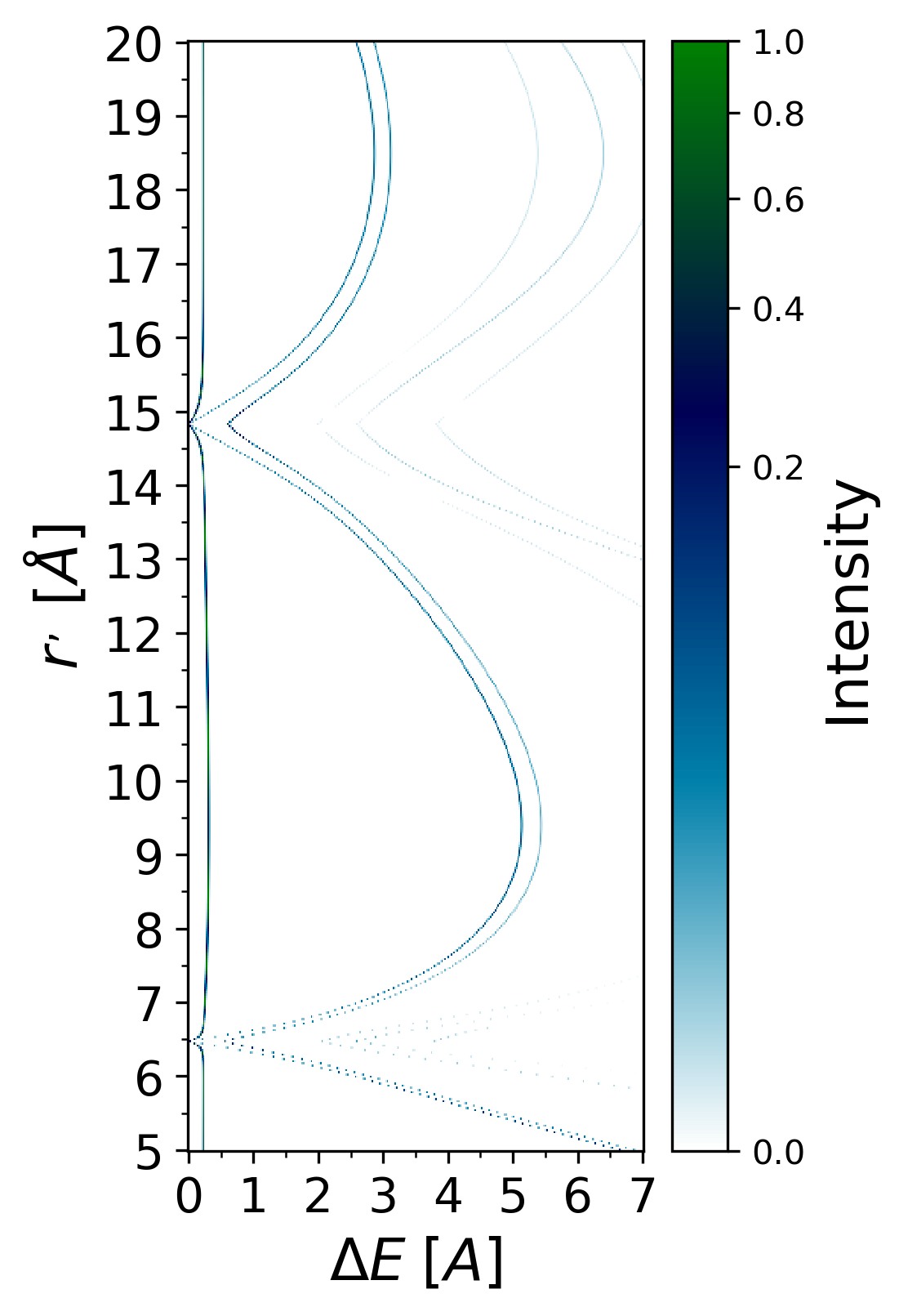}%
        \put(150,900){(a) FM}%
        \end{overpic}
        \begin{overpic}[width=0.45\columnwidth]{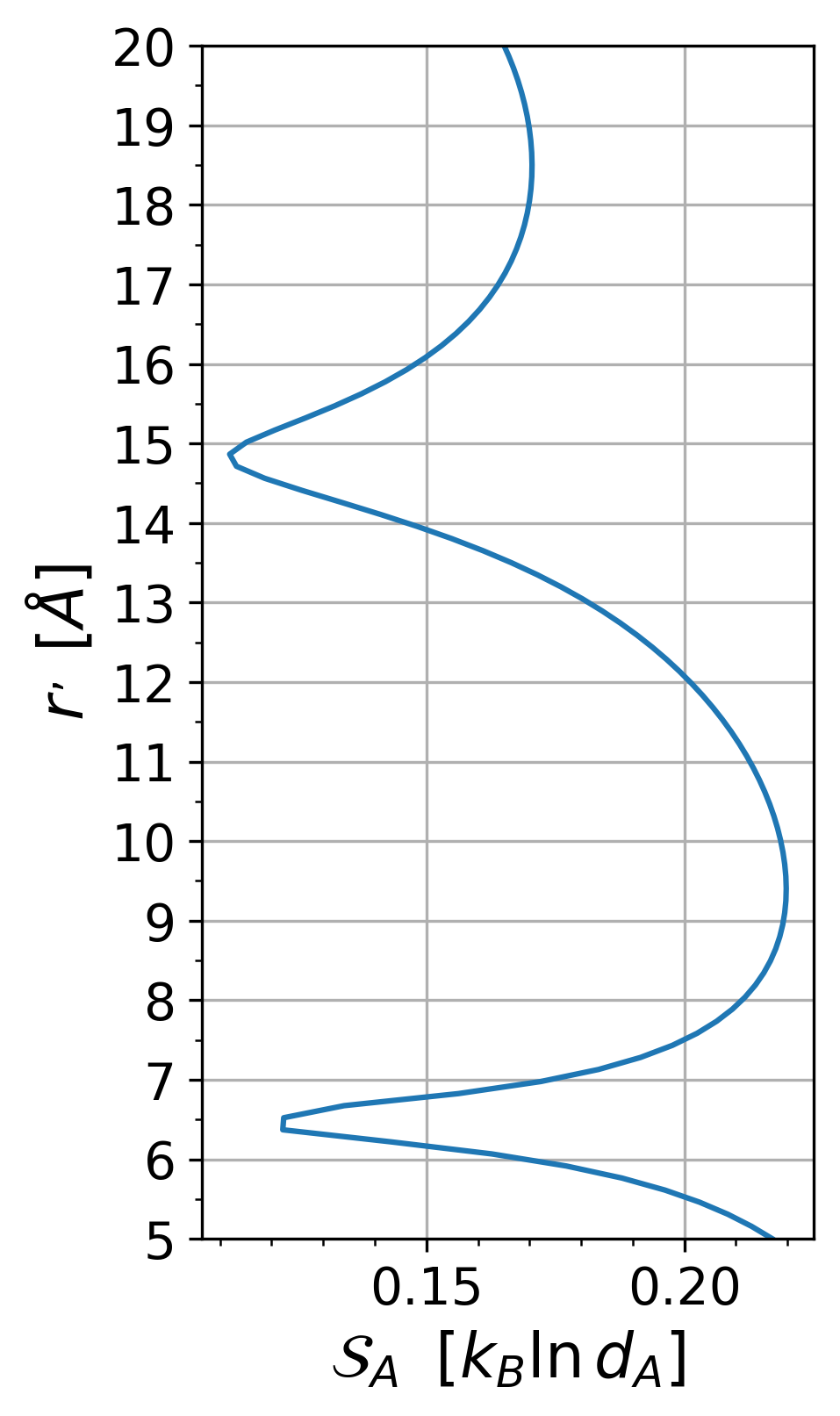}%
        \put(150,900){(b) FM}%
        \end{overpic}
        \begin{overpic}[width=0.52\columnwidth]{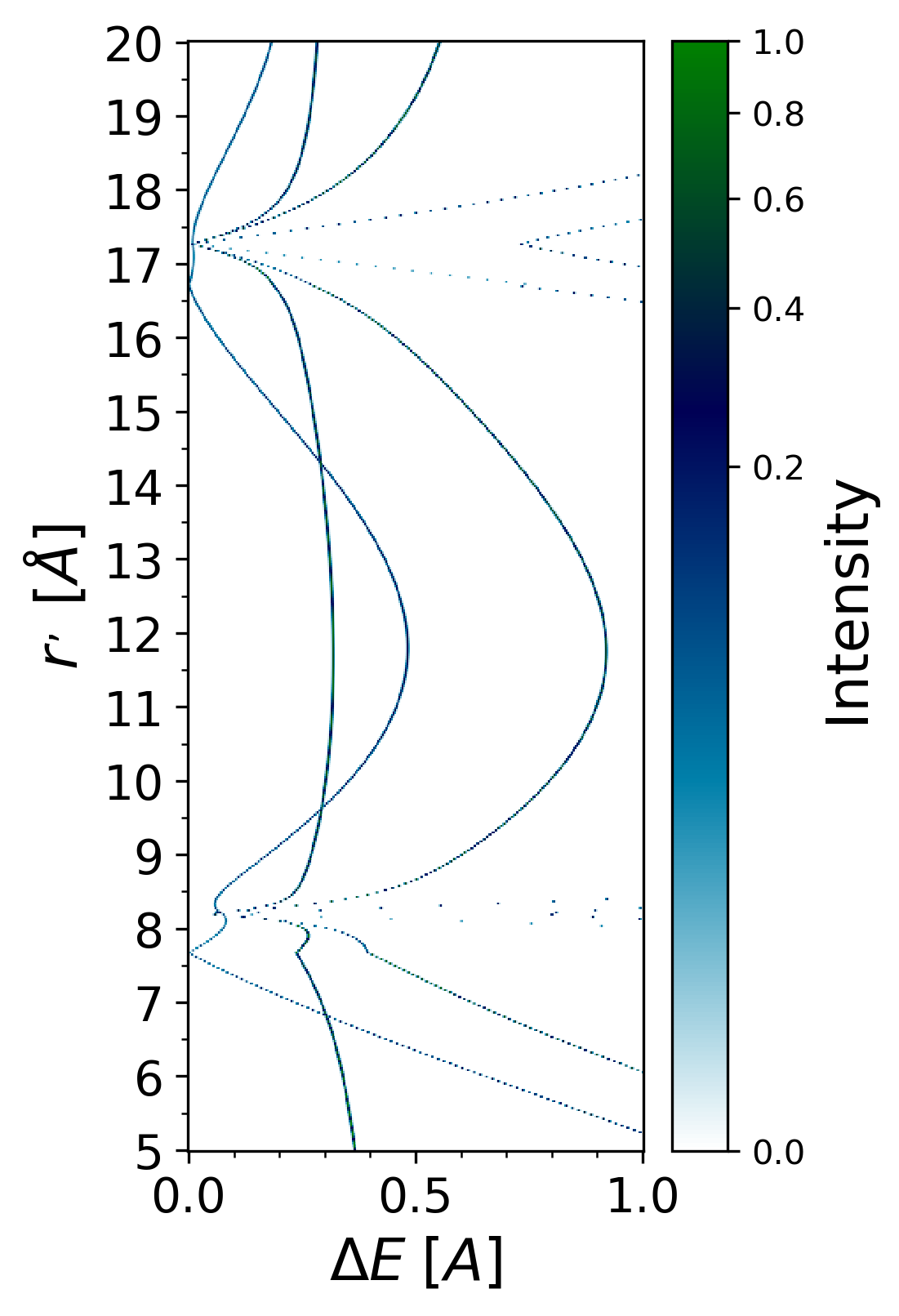}%
        \put(150,900){(c) AFM}%
        \end{overpic}
        \begin{overpic}[width=0.45\columnwidth]{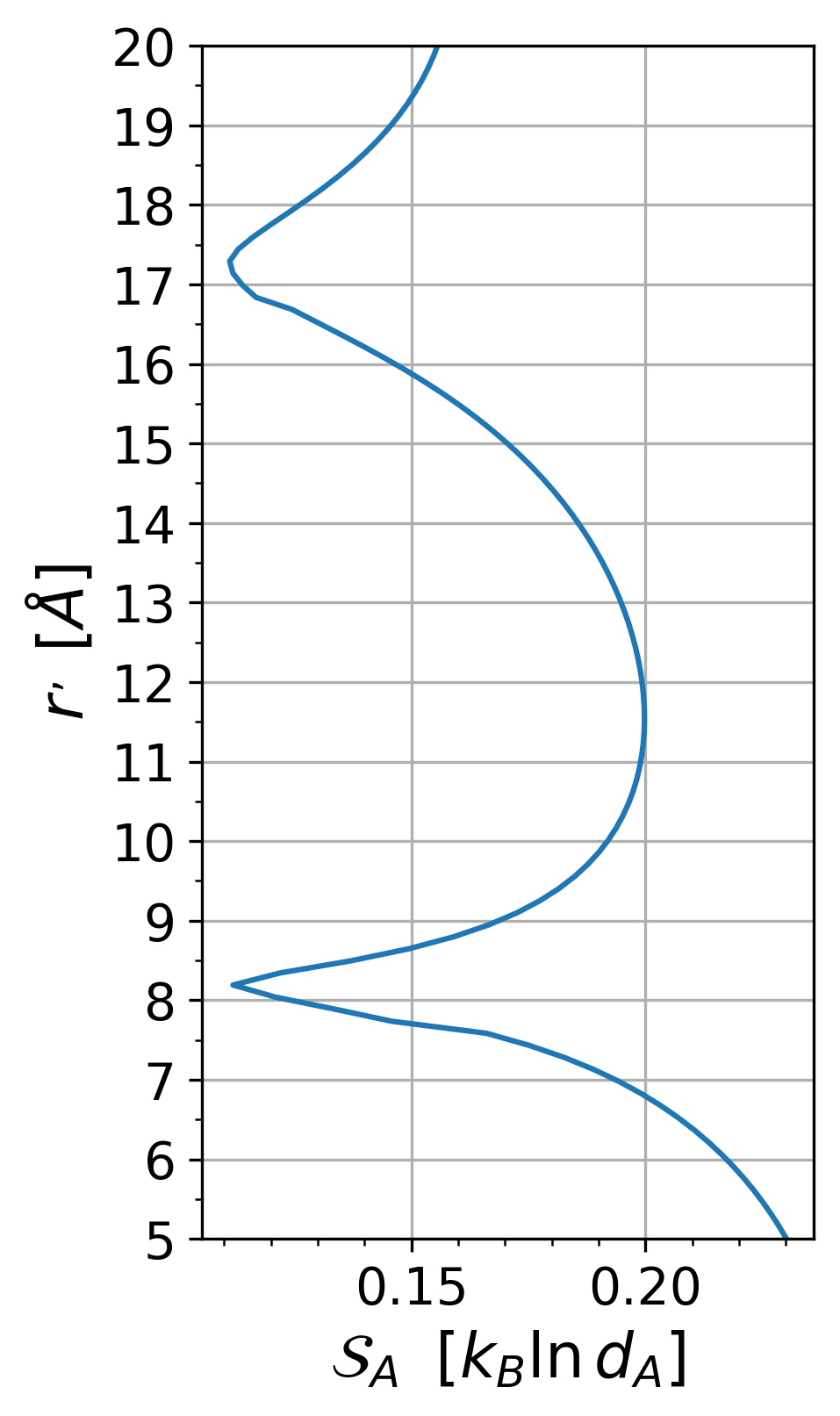}%
        \put(150,900){(d) AFM}%
        \end{overpic}
    \end{center}
\caption{Tunneling spectral (differential conductance) profiles for (a) ferromagnetically coupled and (c) antiferromagnetically coupled triangular clusters as a function of separation of probe-ion and zeroth corner ion at $\theta=0^{\circ}$. Plots are taken for tunneling on the zeroth ion. (b,d) show respective EE plots after probe-ion is traced out. Oscillations and revivals in EE as the probe-ion moves away are due to variations in the RKKY couplings. Notice the different energy range in (a) and (c).} \label{dc_geo}
\end{figure}

Measuring the differential conductance at the zeroth site while moving the probe-ion farther away from the cluster, at $\theta=0^{\circ}$ for example, we can see a correspondence between differential conductance profile peaks (height and energy location), and the EE plots. The tunneling spectral profiles are shown in Fig.\ \ref{dc_geo}(a,c) as functions of $r'$ (with $\theta=0^{\circ}$).

To understand the relative extrema in the corresponding EE plots, Fig.\ \ref{dc_geo}(b,d), we look again at the effective exchange couplings between subsystems in Fig.\ \ref{ee_geo}(b,d). Probe-ion positions where the coupling sum $J_3$ is zero correspond to minima in the EE, while relative maxima/minima in $J_3$ correspond to maxima in the EE. Thus, manipulating the location of the probe ion affects the ground state of the cluster and the respective EE remaining after tracing out the probe. 

%%%%%%%%%%%%%%%%%%%%%%%%%%%%%%%%%%%%%%%%%%%%
\section{\label{Conclusion}Conclusions}
Tunable quantum systems represent a fascinating frontier in condensed matter physics, with their sensitivity to environmental factors offering both challenges and opportunities for material design. Our results demonstrate that finite two-dimensional clusters of lanthanide-based ions, such as Eu$^{2+}$ on metallic substrates, exhibit robust nonlocal correlations driven by magnetic exchange and single-ion anisotropy. These correlations, reflected in the non-vanishing entanglement entropy of the subsystems, can be qualitatively assessed using scanning tunneling microscopy, linking differential conductance to entanglement in a practical and experimentally accessible manner. Our findings are  relevant in light of recent experiments demonstrating  energy conversion in a quantum engine with entangled $^{40}$Ca$^+$ ions, where tuning the degree of entanglement in the system affects the amount of extractable energy \cite{zhang2024}. The design and implementation of systems with well-controlled EE may be useful to further test energy transport protocols as those studied recently in spin chains and molecules \cite{hotta2009, rodriguez-briones2023}.

The dual capability of atomically manipulating system geometry and quantifying entanglement in a single experimental run--whether by varying ion separations or applying external magnetic fields--provides a versatile framework for studying quantum correlations. Our findings establish a theoretical foundation for the interesting relationship between differential conductance and entanglement, paving the way for advances in quantum system fabrication and characterization.

Although our approach lays essential groundwork, further exploration is required to quantify these correlations in increasingly complex geometries and under diverse experimental conditions. We anticipate that this work will stimulate broader efforts in integrating entanglement-based metrics into quantum material design, advancing both theoretical understanding and practical applications in quantum information science.

%%%%%%%%%%%%%%%%%%%%%%%%%%%%%%%%%%%%%%%%%%%%

\begin{acknowledgments}
We acknowledge support from the U.S. Department of Energy, Office of Science, Office of Basic Energy Sciences, Materials Science and Engineering Division. DWF thanks the Nanoscale \& Quantum Phenomena Institute for a student research fellowship.
\end{acknowledgments}

%%%%%%%%%%%%%%%%%%%%%%%%%%%%%%%%%%%%%%%%%%%%
\appendix

\section{The role of SIA}
The role of SIA, especially in terms of competing with the out-of-plane field, is further explored here. Figure \ref{AFM_dcee_halfA_field} shows the tunneling spectral function and corresponding EE plot for an antiferromagnetic cluster system similar to that in Fig.\ \ref{AFM_dcee_field}, but where a weaker SIA is assumed (i.e., only half as strong, $A' = 0.5A = 0.025$ meV). When viewed against Fig.\ \ref{AFM_dcee_field}(a,b), one notices that reducing the SIA value decreases both the energy width of the braiding and field depth in the tunneling spectral function. The smaller field depth is also reflected in the EE. As before, a one-tangle correlates perfectly with the saturation field that fully polarizes the ground state and annihilates any EE in the system. We find that further reducing the SIA follows the same trend. Thus, one can conclude that SIA ensures robustness of EE against competing magnetic fields.
\begin{figure}[h]
    \begin{center}
        \begin{overpic}[width=0.52\columnwidth]{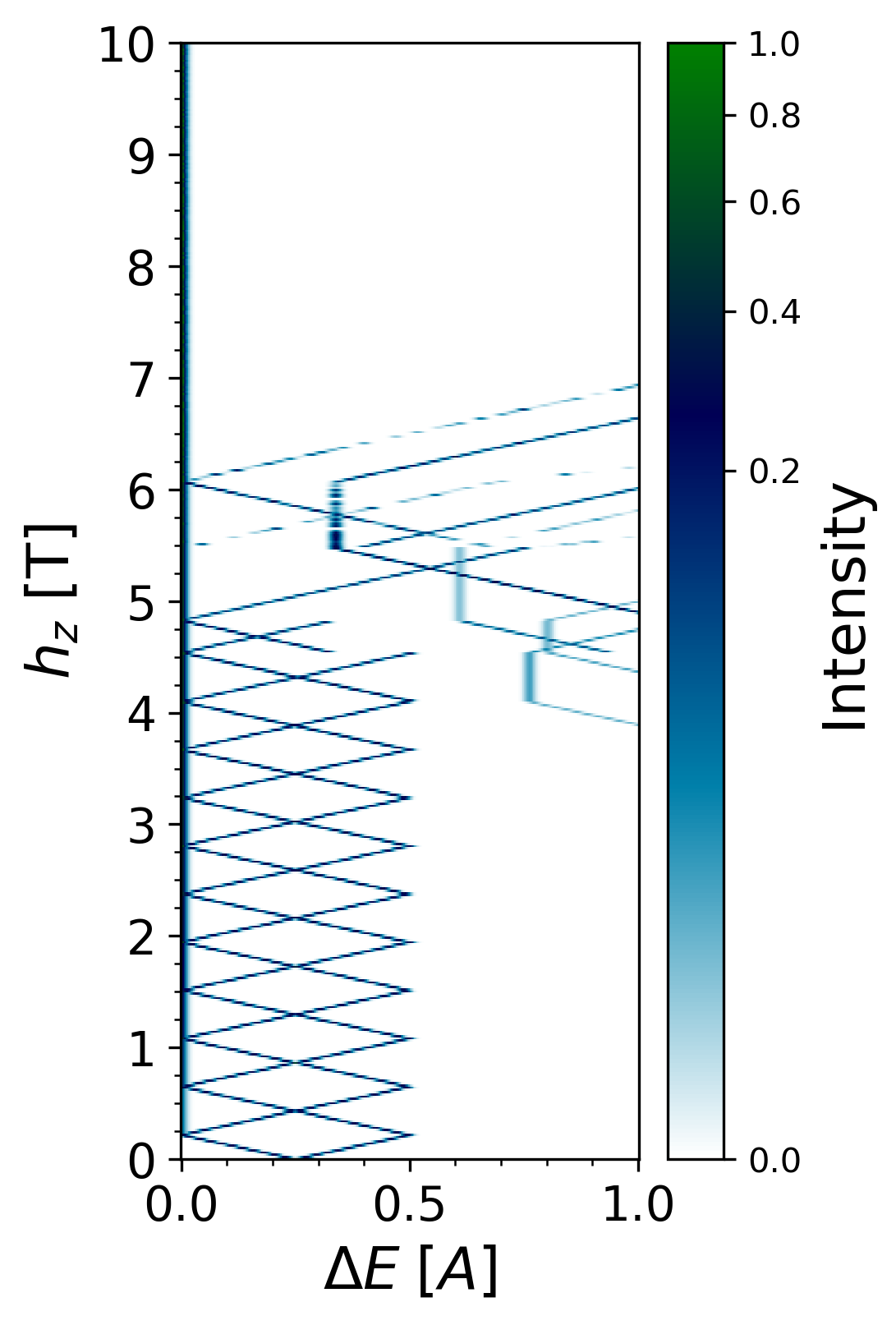}%
        \put(150,900){(a)}%
        \end{overpic}
        \begin{overpic}[width=0.45\columnwidth]{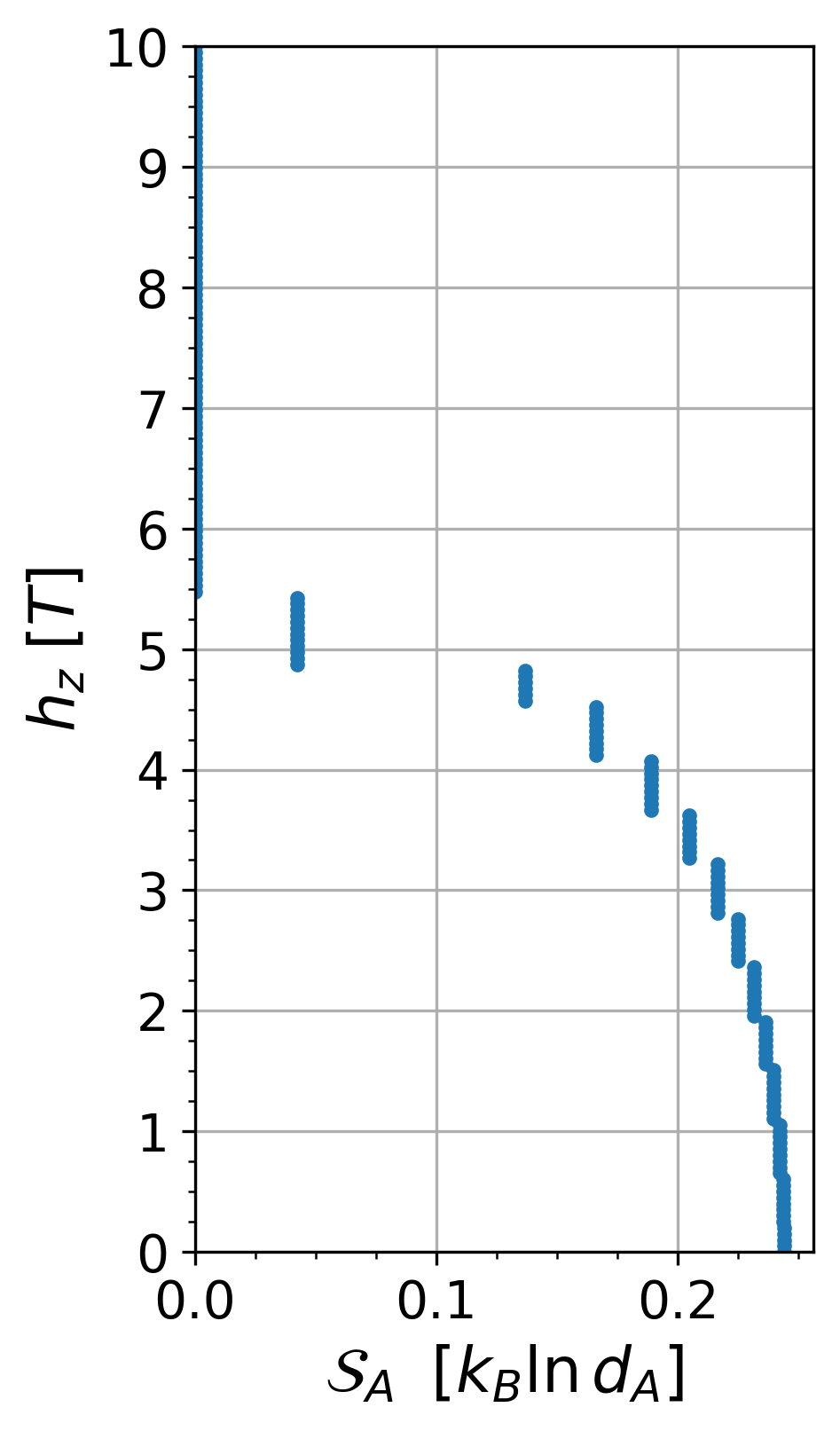}%
        \put(150,900){(b)}%
        \end{overpic}
    \end{center}
\caption{Antiferromagnetic cluster with half-strength SIA, $A' = 0.5A = 0.025$ meV. (a) Color map of the tunneling spectral function (proportional to differential conductance) for the antiferromagnetic triangular cluster and Zeeman field along the $z$ direction. The probe-ion is coupled ferromagnetically to the zeroth ion ($r'\approx 5$ \AA, $\theta = 0^{\circ}$). The tunneling is on the zeroth ion. Energy scaling in panel (a) is the same as in all other figures, where $A = 0.05$ meV. Panel (b), shows EE after probe-ion $\{3\}$ has been traced out. Notice narrower energy and smaller field depth of braiding in (a) impacts non-zero EE range in (b). } \label{AFM_dcee_halfA_field}
\end{figure}

%\section{A little more on appendixes}

%\subsection{\label{app:subsec}A subsection in an appendix}

%\nocite{*}

\bibliography{ee}% Produces the bibliography via BibTeX.

\end{document}